\documentclass[11pt]{article}
\usepackage{amsfonts,amsmath,amsthm,array,amssymb}

\usepackage{graphicx}
\usepackage{picinpar,graphicx}
\usepackage{graphics}
\usepackage{multicol}

\usepackage{amssymb}
\usepackage[utf8x]{inputenc}
\usepackage{wrapfig}
\usepackage{subfigure}
\usepackage{amsmath}
\usepackage[font=footnotesize,labelfont=bf]{caption}
\usepackage{booktabs}
\usepackage{rotating}
\usepackage{epsfig}
\usepackage{grffile}
\usepackage{epstopdf}
\usepackage{verbatim}
\usepackage{url}
\usepackage{mathtools}
\usepackage[cm]{fullpage}
\usepackage{url}
\usepackage{enumerate}
\usepackage{cite}

\usepackage[compact]{titlesec} 
 \titlespacing{\chapter}{0pt}{0pt}{0pt}

\begin{document}
\title{A study of gravity-linked metapopulation models for the spatial spread of dengue fever}
\author{Marta Sarzynska$^{1}$, Oyita Udiani$^{2}$, Na Zhang$^{2}$}
\date{}
\maketitle
\begin{center}
\footnotesize $^{1}$ Mathematical Institute, University of Oxford, Oxford, UK\\
\footnotesize $^{2}$ AMLSS, Arizona State University, Tempe, AZ\\
\end{center}

\begin{abstract} 
{
Metapopulation (multipatch) models are widely used to study the patterns of spatial spread of epidemics. In this paper we study the impact of inter-patch connection weights on the predictions of these models. We contrast arbitrary, uniform link weights with link weights predicted using a gravity model based on patch populations and distance. In a synthetic system with one large driver city and many small follower cities, we show that under uniform link weights, epidemics in the follower regions are perfectly synchronized. In contrast, gravity-based links allow a more realistic, less synchronized distribution of epidemic peaks in the follower regions. We then fit a three-patch metapopulation model to regional dengue fever data from Peru -- a country experiencing yearly, spatially defined epidemics. We use data for 2002-2008 (studying the seasonal disease patterns in the country and the yearly reinfection patterns from jungle to the coast) and 2000-2001 (one large epidemic of a new disease strain across the country). We present numerical results. 

}
\end{abstract}


\section{Introduction}
The spatial spread of disease is a common and well established research topic in mathematical biology, both in academic and policy-making setting \cite{Murray2003}. With the increase in computing power in the last 20 years, modeling the spread of infections is a rapidly developing field, seeing a number of novel, increasingly complex and realistic model types. Model results are increasingly applied in policy-making settings, for example during the 2001 foot-and-mouth disease epidemic in the UK \cite{Sattenspiel2009} and the 2009 influenza A/H1N1 (``swine flu'') pandemic \cite{Boelle2011}.

\subsection{Incorporating space into compartmental models}
\label{section:space}
Compartmental models  -- the most common type of disease model -- are usually used to study the progress of a disease within a single, well-mixed population. After validation using  past data they can be used as generative models to predict likely epidemic outcomes and to study disease properties. However, when looking at a larger scale of disease spread, distribution of individuals across space and patterns of interactions between groups become important influences on the spread and persistence of infections. Large scale models of disease tend to incorporate space in some way. 

The majority of spatial models belong to two classes, either metapopulation (multipatch) models or spatially continuous models using partial differential equations \cite{Sattenspiel2009}. These two types of models deal with different questions -- metapopulation models represent sets of communities joined via a contact network which can (but does not have to be) distance-dependent. These models are well suited to spread of disease between people, who tend to follow predefined transport routes, often on a fast time scale (e.g. plane travel). In contrast, continuous models explicitly model the space in which the disease spreads, often leading to reaction-diffusion equations with traveling wave solutions. This type of models is better suited to modeling animal populations which can be assumed to disperse at random, or human populations before the invention of trains and planes -- they have been used for modeling plague in the middle ages \cite{Murray2003} In this study we will focus on metapopulation models. 

Metapopulation models have been used for studying coupling patterns between populations, spatiotemporal patterns of disease persistence and synchrony between populations, and hierarchical transmission between cities, towns and villages \cite{Xia2004,Viboud2006}. They have also been used on a larger scale to model global spread of pandemics, such as SARS \cite{Hufnagel2004} and H1N1 \cite{Balcan2009}.

In this study we implement a country-wide metapopulation model with realistic connectivity structure obtained from a gravity model -- a type of model that is commonly used to predict different types of connectivity between cities, e.g. transport, trade, and population movement \cite{Barthelemy2011}. We begin by implementing a synthetic model, with one large, central city driving epidemics in 99 small cities around it. We study the influence of inter-patch link weights on metapopulation model predictions by comparing the spatial patterns of disease spread in uniform and gravity-linked models. We then create a deterministic gravity-linked metapopulation model for dengue fever and fit it to data on the local incidence of dengue in Peru between 1994 and 2008. We focus on calibrating the model to the epidemic patterns observed in the different climatic regions of the country using a region-specific, time-dependent vector to host transmission parameter.

\subsection{Gravity models}

Gravity models assume the interaction between two locations is proportional to their importance (e.g. population), but it decays with distance \cite{Barthelemy2011}. Spread of dengue fever between different provinces is largely due to human travel patterns (as the mosquitoes only travel very short distances \cite{Liebman2012}), so a gravity model is expected to be a good fit for the contact matrix of the multipatch model. 

The general form of a gravity model for interaction $P_{ij}$ between locations $i$ and $j$ with populations $n_i$ and $n_j$ respectively, located at distance $d_{ij}$ apart, can be described as 

\begin{equation}
P_{ij} = \theta \frac{n_i^\alpha n_j^\beta}{d_{ij}^\gamma},
\label{eq:gravitymodel}
\end{equation}
where $\alpha$, $\beta$ and $ \gamma$ are parameters that are commonly determined by regression analysis and $\theta$ is a scaling factor used to adjust the scale of the interactions to fit the disease spread data  \cite{Barthelemy2011}.

\subsection{Fitting models to data}

In this paper we will attempt to obtain some of the parameters of the disease model and all the gravity model parameters ($\alpha, \beta, \gamma$ and $\theta$) by fitting model predictions with different parameters to the disease data. Two main methods for fitting models to data are least squares statistic and Pearson chi-squared statistic. Both statistics measure the distance between data points and model predictions at the same time points. 

The least squares statistic is one of the most commonly used goodness of fit statistics. It is defined as the sum of the squared point-by-point distances between the model prediction and the data:

\begin{equation}
S = \sum_{i=1}^{N} (M_i - D_i)^2, \nonumber
\end{equation}
where $N$ is the number of time points, $M_i$ is the model prediction for disease incidence at time point $i$ and $D_i$ is the data for that time point. The best fitting model across many runs with different parameter sets is one that minimizes the least squares statistic. However, one potential issue is that in the least squares method we assume that each data point has the same amount of stochastic variation, which can be an issue for example due to the fact periods of low disease incidence carry relatively large stochasticity. The Pearson chi-squared statistic deals with this issue by weighting the distances and thus taking the stochasticity into account. The Pearson chi-squared statistic is defined as

\begin{equation}
T = \sum_{i=1}^{N} \frac{ (M_i - D_i)^2}{M_i}. \nonumber
\end{equation}
We use it to select a best fitting model in the same fashion as the least squares.

\section{The disease and the data}

\subsection{Dengue fever}

Dengue is a human viral infection prevalent in most tropical countries and carried mainly by the \textit{Aedes aegypti} mosquito. There are four dengue strains (DENV1-4); these viruses cause both dengue fever (DF, usually a self-limiting illness running a typical course without need for treatment) or more dangerous dengue hemorrhagic fever (DHF) and dengue shock syndrome (DSS). Infection with one strain produces lifelong immunity against reinfection with that strain but no long-term protection against the other three DENV types \cite{Nishiura2008}. Subsequent infection with a different strain is usually associated with more severe disease such as DHF or DSS \cite{Guzman2010}. Yearly epidemics of dengue in mos countries are thought to often be facilitated by different strains dominating the disease landscape each year. 

In the Americas, the cancellation of mosquito eradication programs in the 1970s facilitated the re-emergence of dengue fever, and it is currently the most prevalent vector-borne disease \cite{Guzman2003,Chowell2011}. In Peru, DENV-1 was first recorded in 1990 in Iquitos in the Amazon region \cite{Chowell2008}. The American genotype DENV-2 was the second to invade and it is thought to be the driver of a large dengue epidemic focused on Iquitos in 1995-1996 \cite{Chowell2008}. The large countrywide epidemic of 2000-2001 was the first in which all four dengue serotypes co-circulated \cite{Chowell2008}, and the country has been experiencing yearly outbreaks since.

In this project we focus on two periods. The first is 2000-2001, during the large countrywide epidemic that was the first exposure of Peru to DENV 3-4 and thus can be treated as a well-mixed, single epidemic. The second period of interest is post-2002, when Perus started experiencing yearly epidemics.

Dengue transmission occurs principally through a bite of infected mosquitoes. Female mosquitoes acquire DENV by biting infected humans, and they become infective after an extrinsic incubation period of 8-12 days (it takes longer the lower the temperature) \cite{Focks1995}. They can then transmit DENV for the rest of their life (the mosquito lifespan is approximately 1 month) \cite{CDC}. Following infection, the incubation period is typically 4-7 days (range of 3-14 days) \cite{Guzman2010}. The \emph serial interval, defined as the time between onset of symptoms in successive cases in a chain of transmission, is estimated at 15-17 days \cite{Aldstadt2012,Fine2003}. 

No effective antiviral agents to treat dengue infection are currently available; supportive treatment with particular emphasis on careful fluid management is the prevailing approach \cite{CDC}. Vector control, through chemical or biological targeting of mosquitoes and removal of their breeding sites, is the main method of dengue prevention \cite{Simmons2012}. 
Despite many years of effort, it has been difficult to develop a vaccine for dengue, due to the fact any vaccine would have to be protecting against all four strains, otherwise it would expose patients to increased risk of DHF \cite{CDC}. Epidemiological models that can predict the likelihood of an outbreak given prior infection patterns can inform policy makers in deciding where and when to apply controls.

\subsection{Peru}
\label{section:countries}
Peru is located on the Pacific coast of South America, between $3^\circ$ S and $18 ^\circ$ S. The population of about 29 million is heterogeneously distributed throughout a surface area of 1,285,220 km$^2$, composed of a western coastal plain, the Andes mountains in the center, and the Amazon jungle in the east. The weather varies from dry in the coast and tropical in the Amazon to cold in the Andes. These heterogeneities influence disease transmission, as mountains form a natural barrier to the spread of the mosquitoes and hence dengue fever. The jungle forms a reservoir of endemic disease, from which occasionally it spreads across the country in an epidemic \cite{Chowell2008}. 

Peru is divided in 25 administrative regions, which are further subdivided into 195 provinces. Our data is gathered at the province level. Each province is classified as ``Mountain'', ``Coast'', or ``Jungle'', with a further classification of the first two into ``northern'', ``central'', and ``southern''. We obtained census data from the National Institute of Statistics and Informatics of Peru, including population size per province in 1994 \cite{PeruStats}. We calculated estimates of population density per province (people/km$^2$), and they range from a mean of 22.3 people/km$^2$ in the mountains, 12.38 in the jungle, and 172 in the coastal areas. 

\subsection{The data}
The dengue data set consists of weekly regional time series recording the number of probable and confirmed dengue cases, as recorded by the Peruvian Health Ministry's General Office of Epidemiology. It contains data for 79 provinces over 780 weeks between 1994 and 2008; the remaining 116 provinces did not record any cases during that period. A total of 86,631 dengue cases were reported, mainly in jungle and coastal regions (47\% and 49\%, respectively) and only 4\% in the mountains. 



We downloaded climate data for weather stations nearest each of the climatic regions from the National Climate Data Center (NCDC) of the National Oceanic and Atmospheric Administration (NOAA), via the weatherunderground.com website. The data includes daily maximum, minimum and average temperature, dew, humidity, wind, pressure, and precipitation. Previous research showed that the most influential weather factors on the development of dengue are minimum temperature and humidity, as they affect the external incubation period of dengue in mosquitoes and the larval development \cite{Focks1995,Chowell2008}. In this model we will focus on the effects of the most important climatic variable -- minimum temperature. 

\section{The model}

We construct a metapopulation model with separate vector and host populations in each patch, inspired by the two-patch model that was fitted to the same data set  by Torre et al. \cite{Torre2009}. They focused on the influence of transportation between patches on the temporal sequence of disease spread between jungle and coast regions, and showing the influence of transportation and basic reproduction number $R_0$ on the relative timing of epidemic peaks. Their model mostly focused on qualitative features of the epidemics, such as the location of the peaks and the distance between them. However, the numbers of disease cases it predicted was overestimated as it predicted infection of up to 20\% of the population, whereas the highest proportion of infected individuals per province in our data set is 1.2\%, suggesting we might need to revise the parameters taken form that model. 

We define a deterministic model for the disease in each patch, its structure inspired by Torre et al. \cite{Torre2009}, much enlarged and with the addition of gravity links between patches. This is the first time to our knowledge that such explicit formulation of human transportation has been incorporated into a vector-host model. We assume only human travel contributes to the spread of disease, as the mosquitoes only travel a very limited distance  \cite{Liebman2012}. The model for patch $i$ out of $n$ patches: vector (${S}_{vi}, {E}_{vi}, {I}_{vi}$) and host (${S}_{hi}, {E}_{hi}, {I}_{hi}, {R}_{hi}$) is shown below and the meanings and units for all the variables and parameters are described in Table \ref{table:modelparameters}.
\begin{align} \nonumber
\dot{S}_{vi} &= \mu_{vi} N_{vi} - \beta_{vi} (t) \left (\sum_{i=1}^{n}{P_{ij} \frac{I_{hi}}{N_{hi}}} \right )S_{vi} - \mu_{vi}S_{vi}\\ \nonumber
\dot{E}_{vi} &= \beta_{vi} (t) \left (\sum_{i=1}^{n}{P_{ij} \frac{I_{hi}}{N_{hi}}} \right )S_{vi} - \mu_{vi}E_{vi} - \kappa E_{vi}\\ \nonumber
\dot{I}_{vi} &=  \kappa E_{vi} - \mu_{vi}I_{vi}\\ \nonumber
\dot{S}_{hi} &= \mu_{hi} N_{hi} - \beta_{hi} P_{ii} \frac{I_{vi}}{N_{vi}} S_{hi} - \mu_{hi}S_{hi}\\ \nonumber
\dot{E}_{hi} &=  \beta_{hi} P_{ii} \frac{I_{vi}}{N_{vi}} S_{hi} -\lambda E_{hi} - \mu_{hi}E_{hi}\\ \nonumber
\dot{I}_{hi} &= \lambda E_{hi} -\delta I_{hi} - \mu_{hi}I_{hi}\\ \nonumber
\dot{R}_{hi} &= \delta I_{hi} - \mu_{hi}R_{hi},\\ \nonumber
\end{align}
where  $\beta_{vi} (t) =  \beta_0 + \epsilon \sin(\frac{2 \pi t + \phi}{365})$,  $\beta_0, \epsilon, \phi$ are transmission fitting parameters, and $P_{ij} = \theta \frac{n_i^\alpha n_j^\beta}{d_{ij}^\gamma}$ is taken from a gravity model, where $\theta, \alpha, \beta$ and $\gamma$ are additional gravity model parameters described below. 

We take the majority of disease related parameter values from Ref.  \cite{Torre2009} and \cite{Chowell2008}, which both fitted compartmental models to the Peru data. The transportation between the patches $i$ and $j$ is defined as a coupling parameter $P_{ij}$, which we will define based on the populations of patches $i$ and $j$ and the distance between them using the gravity model and data fitting. We also carry out numerical simulations and data fitting in single isolated patches to obtain ranges for the base line and the periodic form of the transmission parameter $\beta_{vi}(t)$ for each patch. The parameters and values are in Table \ref{table:modelparameters}.

\begin{table}[!htbp]
\label{table:modelparameters}
\small
\centering
\begin{tabular}{lll}
\hline
Variable&Value & Meaning [units]\\
\hline
$S_{vi}(t)$ & & Susceptible vectors in patch $i$  [vectors] \\
$E_{vi}(t)$ & & Exposed vectors in patch $i$  [vectors] \\
$I_{vi}(t) $& & Infected vectors in patch $i$  [vectors]  \\
$N_{vi}(t) $& $3 \times N_{hi}$ & Total vector population in patch $i$  [vectors]\\
\hline
$S_{hi}(t) $&  &Susceptible hosts in patch $i$  [individuals] \\
$E_{hi}(t) $&  &Exposed hosts in patch $i$  [individuals] \\
$I_{hi}(t) $&  &Infected hosts in patch $i$  [individuals] \\
$R_{hi}(t) $&  &Recovered hosts in patch $i$  [individuals] \\
$N_{hi}(t) $ &  From data  \cite{PeruStats} &Total host population in patch $i$  [individuals] \\
\hline
Parameter&Value & Meaning [units]\\
\hline
$P_{ij}$& gravity model & Probability person from patch $i$ visiting patch $j$ [dimensionless]\\
$\beta_{vi}$&seasonal, varies by patch& Per capita transmission rate from host to vector in patch $i$ [1/days]\\
$\beta_{hi}$& fitted -- rand (0.2,0.5) &Per capita transmission rate from vector to host in patch $i$ [1/days]\\
$\lambda$& $1/5.5$  &Host per capita infected rate  [1/days]\\
$\delta$& $1/4$&Host per capita recovery rate  [1/days] \\
$\kappa$&$1/5.5$&Vector per capita infected rate  [1/days]\\
$\mu$&$1/10.5$&Vector per capita birth/death rate [1/days] \\ 
\hline
\label{table:modelparameters}
\end{tabular}
\caption{The variables and parameters of the multipatch model. }
\label{table:modelparameters}
\end{table}

\section{The synthetic metapopulation model}

\begin{figure} [h!tbp]
\centering
f\hfill(a)   \includegraphics[width=0.45\linewidth]{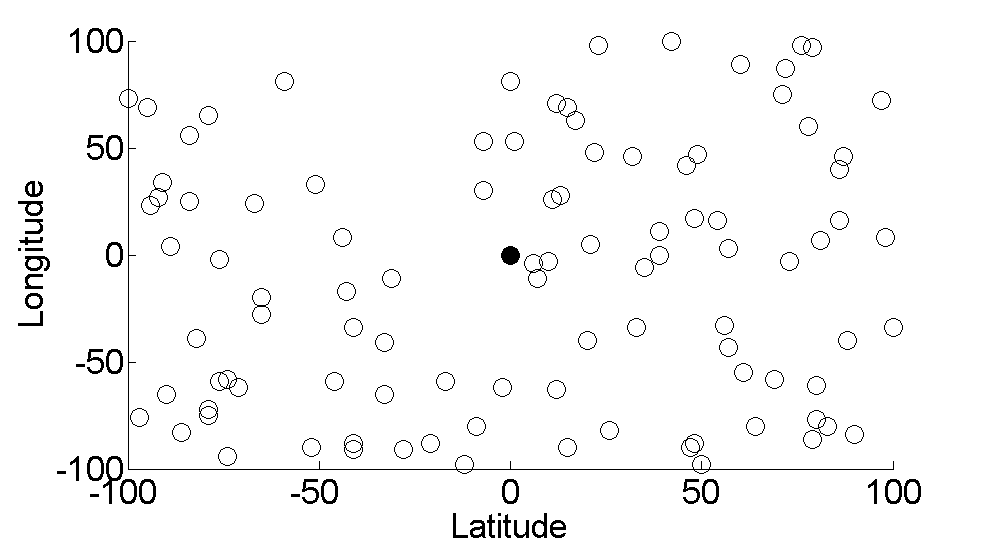}
\hfill(b)  \includegraphics[width=0.45\linewidth]{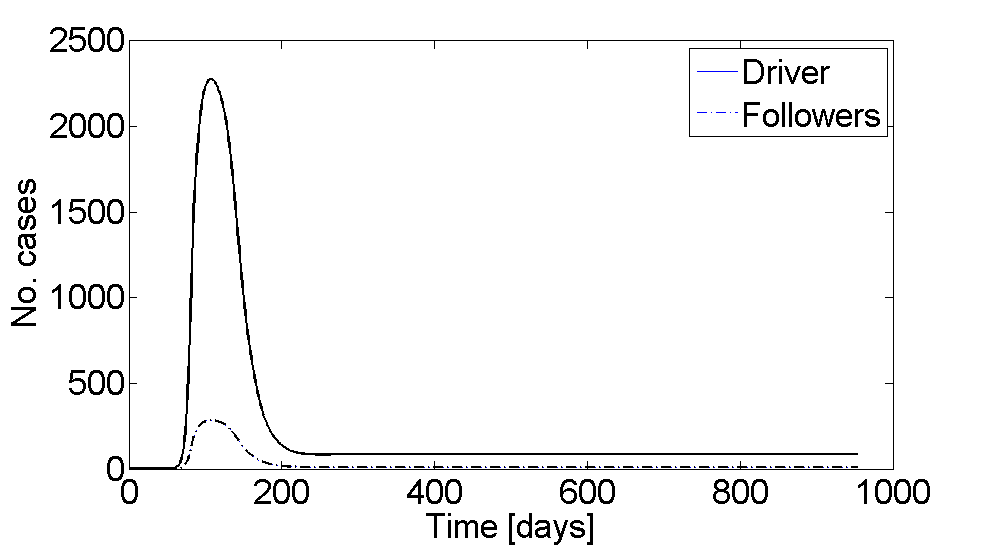}\\
\hfill(c)\includegraphics[width=0.45\linewidth]{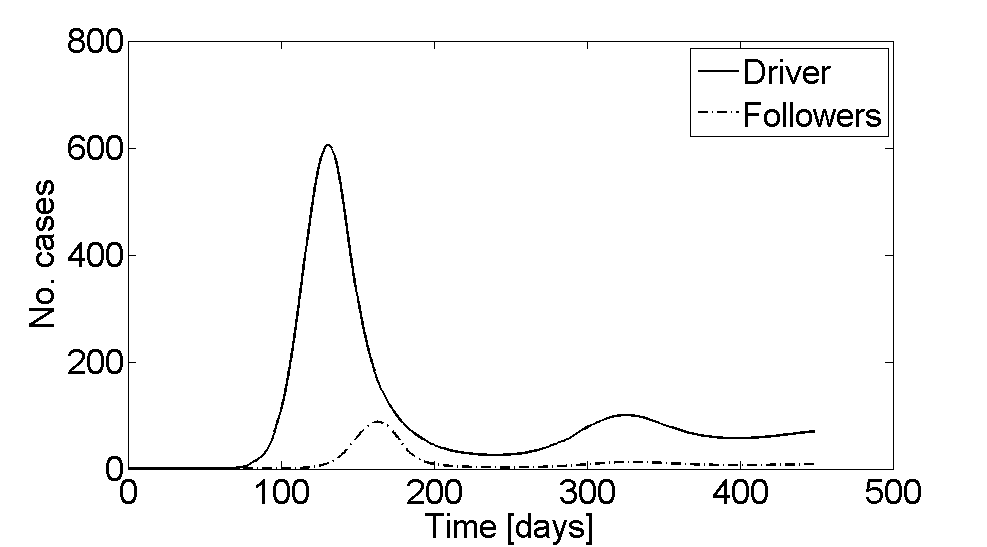}
\hfill(d)\includegraphics[width=0.45\linewidth]{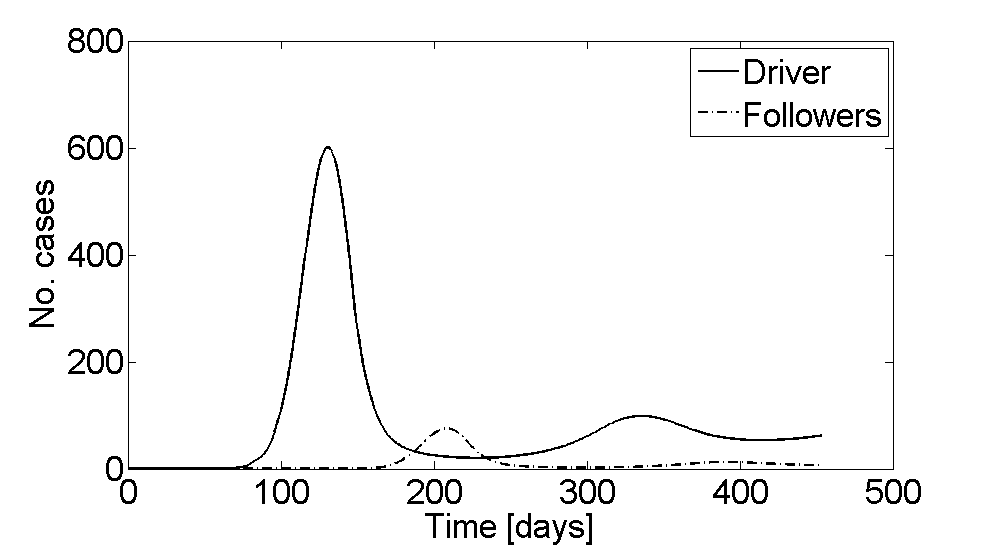}
\caption{The  synthetic multipatch model with uniform links: (a) locations of the cities:  Filled -- driver -- city 1 (population: 8 000 000), circles: follower cities (population 100 000), (b-d) the epidemic curves for connectivity $P(i,j)$, (b) 0.01, (c) 0.001 and (d) $10^-5$. }
\label{fig:nograv} 
 \end{figure}

We first study the behavior of the gravity metapopulation model in a theoretical setting similar to the setting we will be working with for the Peruvian data. We create an artificial set of cities with known distances and populations and study the behavior of the model depending on link weights. The cities are placed in a 200 $\times$ 200 space with coordinates ranging between -100 and 100. We place one large city with population of 8 million in the center at location (0,0) (referred to as city 1). We place 99 small cities with populations of 100 000 people each at random locations on the grid. Figure \ref{fig:nograv}a shows the locations of the cities.  

We seed the epidemic with one infected individual in city 1 (the \emph{driving} city). We compare the behavior of two multipatch models with different link types between cities, and we focus on the patterns of infection in the \emph{follower} cities. In the first model, we give all the links the same weight. In the second model, the link weight is predicted by a gravity model, depending on populations of the cities and the distance between them. 

For simplicity, we set the transmission parameter $\beta_{vi}$ to 0.3, and we have no seasonality in the model. All the other disease related parameters are as previously reported. We study the results from the simple model in Figure \ref{fig:nograv}. At 10$\%$ connectivity, the epidemic waves in the driver and follower cities almost coincide. The delay between the driver and follower cities increases with decrease in connectivity.

\begin{figure} [h!tbp]
\centering
\hfill(a)  \includegraphics[width=0.28\linewidth]{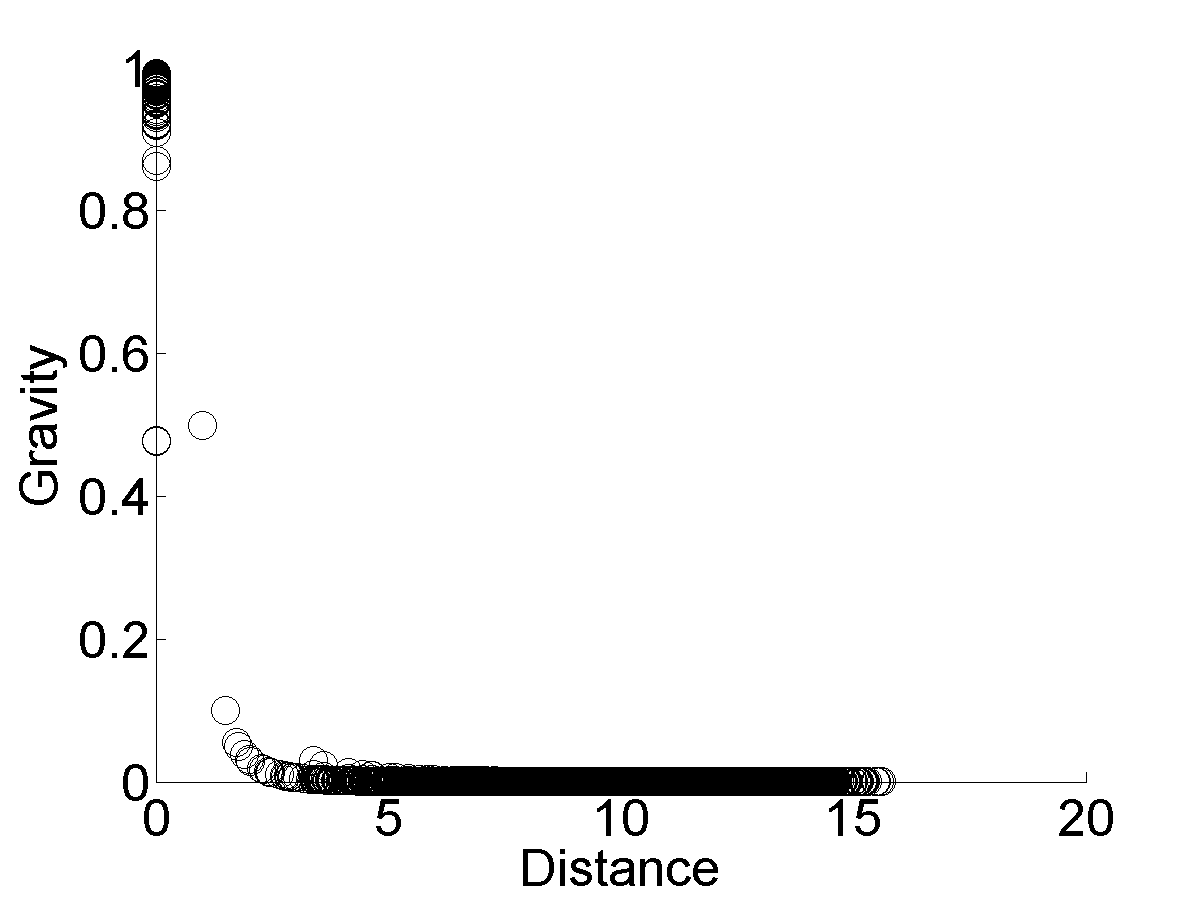}
\hfill(b)  \includegraphics[width=0.28\linewidth]{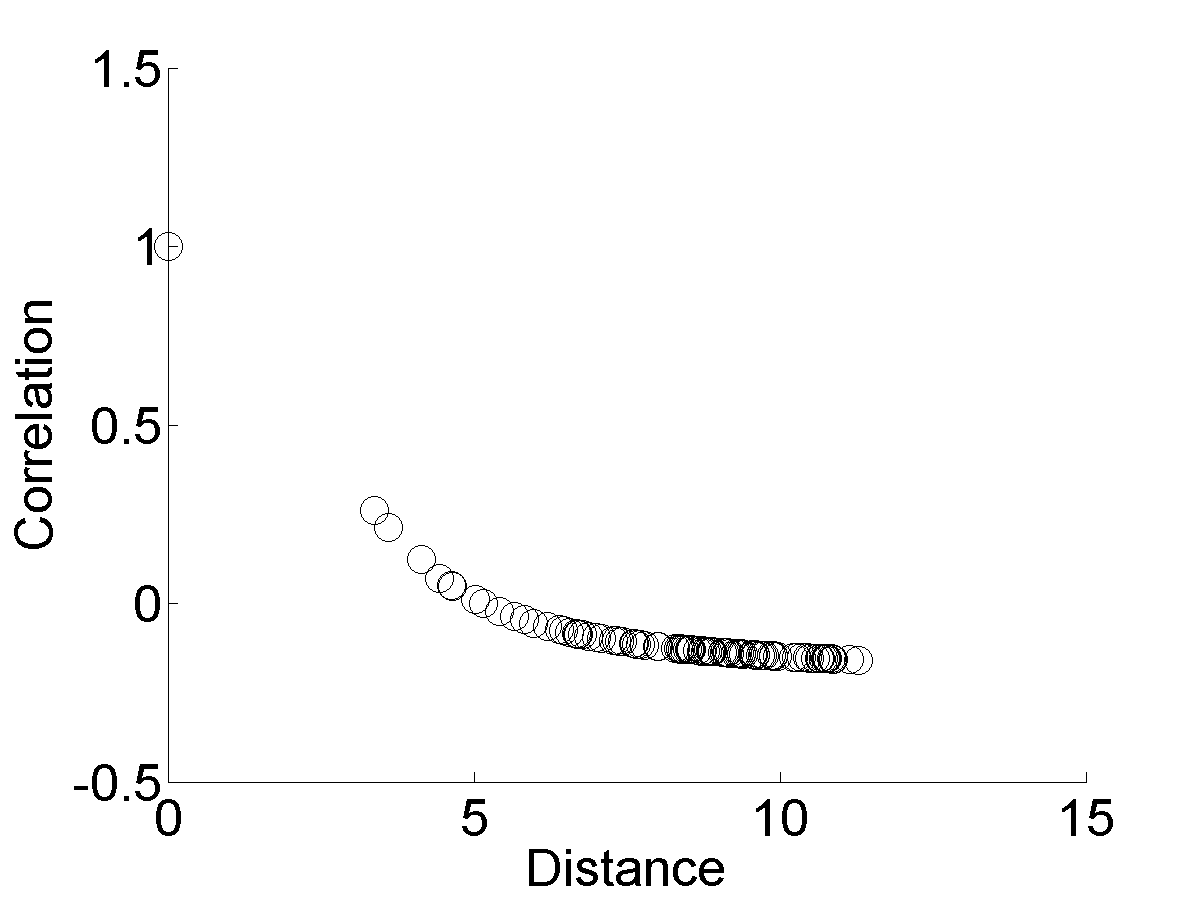}
\hfill(c)  \includegraphics[width=0.28\linewidth]{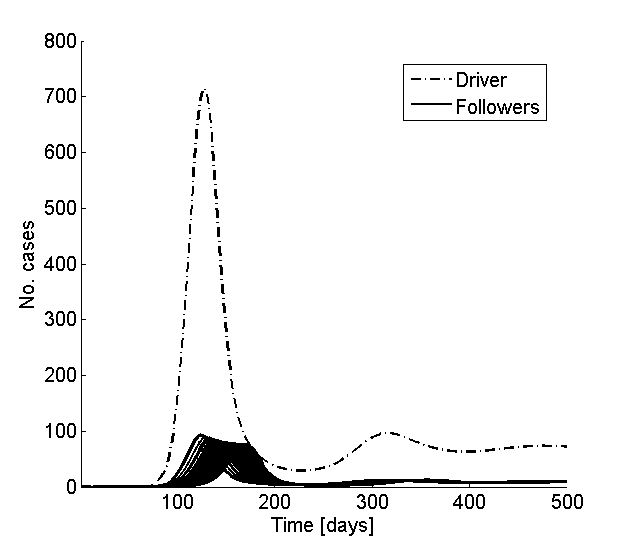}
\caption{The  synthetic multipatch model: (a) the gravity model prediction based on distance, (b) the correlation in the cities' epidemic curve against the epidemic curve of city 1, (c) the epidemic curves. Parameters:  $\alpha=\beta=1$, $\gamma=2$, $\theta=1$.}
\label{fig:grav} 
 \end{figure}

We then study the behavior of the multipatch model with gravity-based links. 
First, we set $\alpha=\beta=1$, $\gamma=2$ and $\theta=0.5$, and we plot the gravity model prediction $G_{ij}$  for the link weight (Figure \ref{fig:grav}a), and the degree of correlation between follower cities and the large city 1 (time series correlation over the whole time) as a function of their distance (Figure \ref{fig:grav}b) -- both are observed to decrease with distance. In Figure \ref{fig:grav}c we show the epidemic curve and notice the different peak locations for the follower cities.

\begin{figure} [h!tbp]
\centering
\hfill(a)   \includegraphics[width=0.45\linewidth]{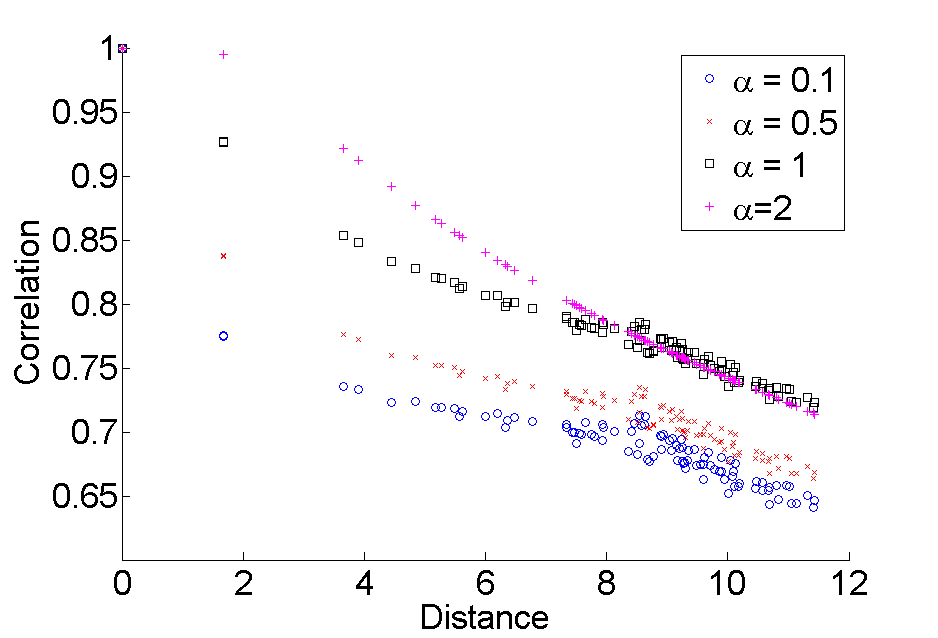}\hfill 
\hfill(b)  \includegraphics[width=0.45\linewidth]{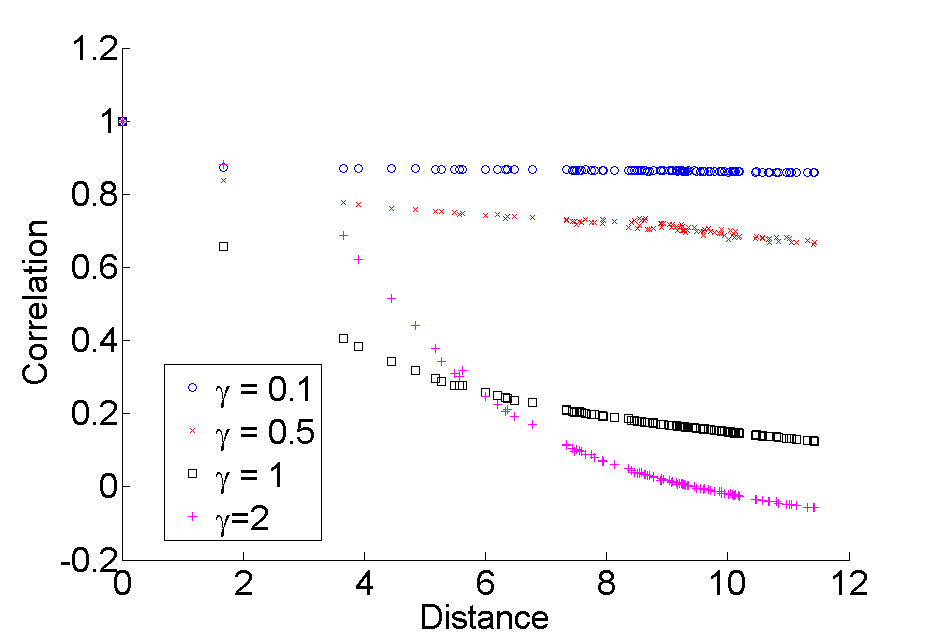}\\
(c)  \includegraphics[width=0.45\linewidth]{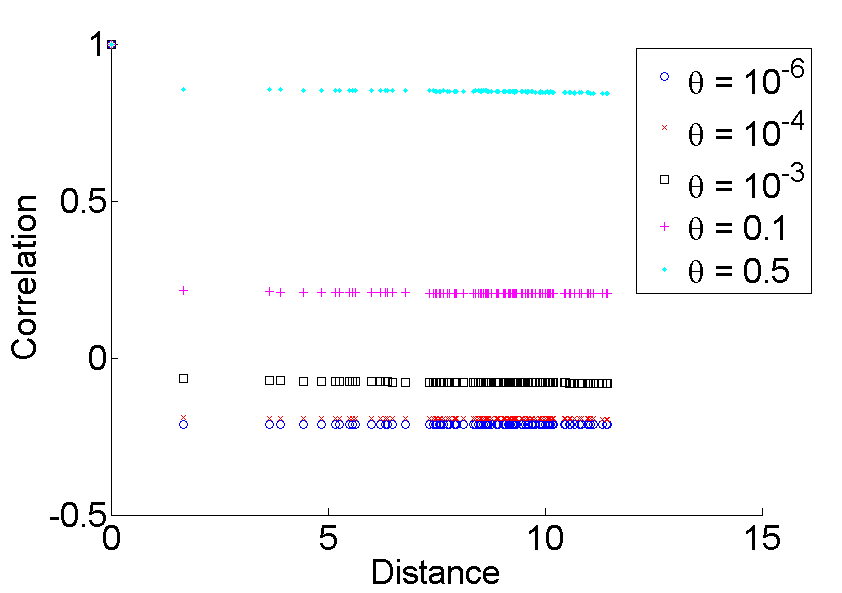}
\caption{The changes in dependence of correlation of all the cities with city 1 on distance as we vary parameters in the synthetic multipatch model: (a) varying the scaling parameter $\alpha$ keeping $\beta = 1, \gamma = 0.5, \theta = 0.5$, (b) varying $\gamma$ keeping $\alpha = 0.5, \beta = 1, \theta = 0.5$ and (c) varying $\theta$ with keeping $\alpha = 0.5, \beta = 1, \theta = 0.5$. }
\label{fig:gravsynthvariation} 
 \end{figure}

We study the changes in performance of the gravity metapopulation model at a variety of parameters: $\alpha, \gamma$ and $\theta$ taken from (0.1, 0.5, 1, 2). For simplicity, we keep $\beta = 1$, meaning that the population of the driver city is more important than the follower city. In Figure \ref{fig:gravsynthvariation} we vary the parameters and observe the changes in the plot of correlation of each of the the follower cities with the driver city 1 against their pairwise distance. We observe an increase in overall correlation as $\alpha$ increases (Figure \ref{fig:gravsynthvariation}a), and an increase in overall correlation as $\theta$ increases (Figure \ref{fig:gravsynthvariation}c). The shape of the correlation curve changes as $\gamma$ is varied, with higher $\gamma$ introducing a stronger distance-dependent decay in correlation (Figure \ref{fig:gravsynthvariation}b). 

These studies show us that the gravity model can be an approximation for transport links between patches. The strength of the link decreases with distance, as does the correlation of disease patterns. Moreover, only the gravity multipatch model enables us to observe staggered delay in epidemic peaks across a range of small (follower) cities.

\section{Setting up the dengue model for Peru}
 In this section we describe the considerations and calculations performed in setting up the model.
\subsection{Choice of patches}
\begin{figure} [h!tbp]
\centering
 \includegraphics[width=0.7\linewidth]{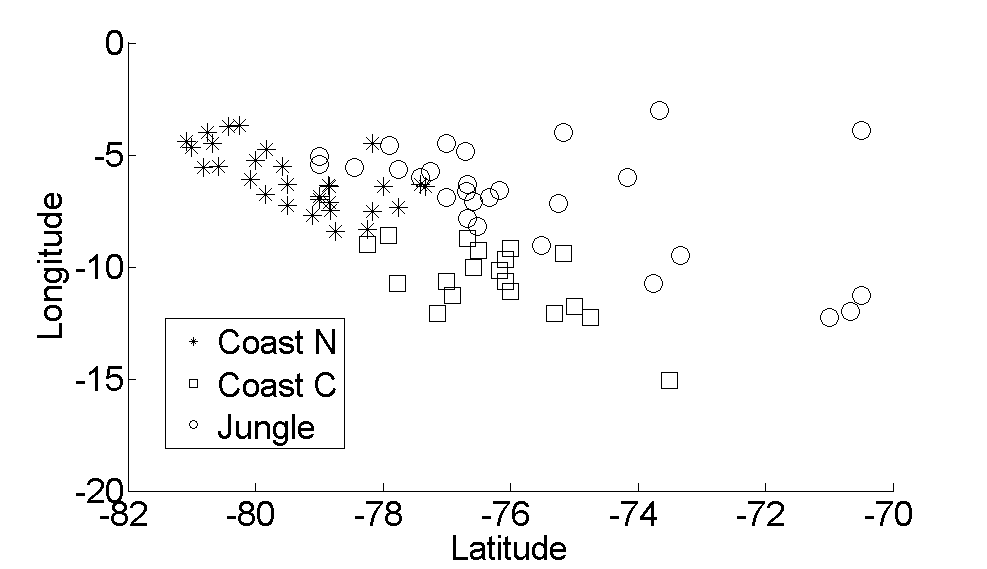}
\caption{The provinces belonging to each of the three patches (o), and the patch centers (*) }
\label{fig:peru_patches} 
 \end{figure}

\begin{figure} [h!tbp]
\centering
\includegraphics[width=0.7\linewidth]{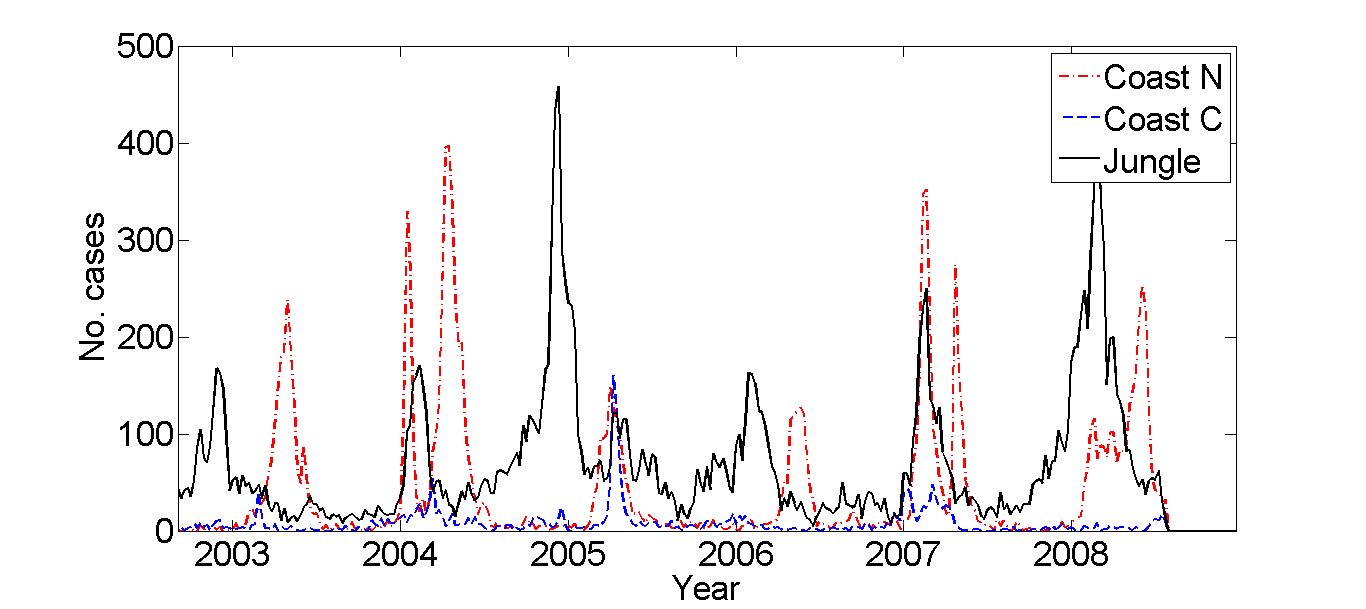}\\
\caption{The number of dengue cases after 2002 in the whole country, and the three patches. }
\label{fig:peru_patchesdata} 
 \end{figure}

The data consists of 79 provinces. Modeling each province separately would potentially give us the most insight into the influence of transport patterns on disease spread, but the fact that not every province experiences an epidemic every year is a potential issue for fitting models to data. 
To simplify the model and avoid irregular data we decided to group the original 7 climate patches into three patches: northern coast, central coast and jungle. Their respective populations in 1994 were 7.6,  10.5 and  2.8 million. The central coast region contains the Peruvian capital, Lima, which holds a large proportion of the population and is an important business and travel hub for the country. The patterns of disease spread differ between the patches: the coastal regions experience higher yearly variation in the number of cases, while the baseline infection rate in winter is much higher in the jungle. It is thought the jungle patch reinfects the coastal regions every spring, and there is a 6 week delay in epidemic peaks between these regions (Figure \ref{fig:peru_patchesdata} ).
Figure \ref{fig:peru_patches} shows the locations of the provinces that belong to each of the 3 patches. Patch centers (required for the gravity model) were calculated as the mean latitude and longitude of the provinces belonging to a patch. Figure \ref{fig:peru_patchesdata} 
shows the patterns of dengue fever infections in the whole country and the three patches, in the selected seasonal epidemic period between 2002 and 2008.

\subsection{Climate data}
\begin{figure} [h!tbp]
\centering
 \includegraphics[width=\linewidth]{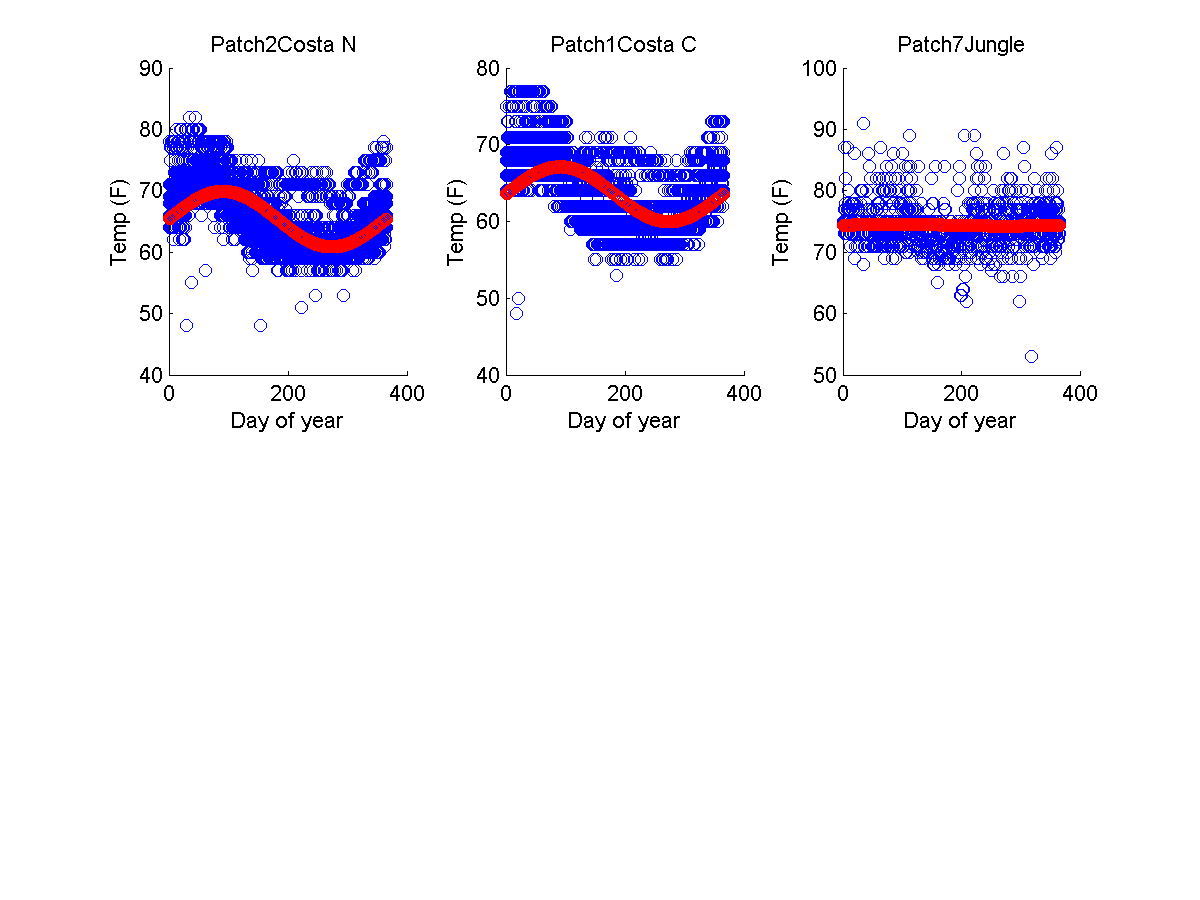}
\caption{The minimum daily temperature across the year in the three patches (blue) and the best sinusoidal fit (red). }
\label{fig:peru_climate3} 
 \end{figure}

We used the minimum temperature data from the weatherunderground.com website for the northern coast, central coast and jungle regions. The three patches have quite different climates. Both coastal regions have large yearly variation in temperature, which is most pronounced in the central coast region. In contrast, the jungle only experiences small seasonal variation in temperature (see Figure \ref{fig:peru_climate3}). This is the reason for dengue fever endemicity in the jungle and lack of it on the coast. 
We used MATLAB's  lsqcurvefit function perform nonlinear regression on the minimum temperature data in order to judge the levels of yearly variation. We fitted a function of the form $T(t) = T_0 + \epsilon \sin(\frac{2 \pi t}{365})$, and the result is overlaid on the data in Figure \ref{fig:peru_climate3}. The coefficients and the proportion of the variation to the mean temperature for each of the regions can be found in Table \ref{table:tempfitting}.
\begin{table}[!htbp]
\label{table:tempfitting}
\small
\centering
\begin{tabular}{lllllll}
Parameter & Coast N& Coast C &   Jungle\\
\hline
$T_0$ [F] &   63.5454   &65.3771 & 74.3880\\
$\epsilon$&    3.5680    &4.5169 &  0.1353\\
$\%$ variation & 5.14 & 6.91 & 0.18\\
\label{table:tempfitting}
\end{tabular}
\caption{The parameters fitting the climate data for the three patches. }
\label{table:tempfitting}
\end{table}

Since minimum temperature is considered the main factor influencing dengue transmission, this gives us an indication of the scale of variation in the transmission parameter $\beta_{vi}(t)$. Basically, when minimum temperature is too low for mosquitoes to transmit the disease, $\beta_{vi}(t)=0$. We considered attempting to link the parameter to temperature directly -- it has been done by Chowell et al. \cite{Chowell2008}. However, incorporating temperature effects directly would require incorporating larval stages of mosquito life cycle, and a variable length of the external incubation period, which would complicate our model beyond the scope of this study. We thus only incorporated the qualitative insight to the shape of the function into our definition of the form of $\beta_{vi}(t)$.

\subsection{Non-linked three-patch model: fitting $\beta_{vi}$}

\begin{figure} [h!tbp]
\centering
\hfill (a)  \includegraphics[width=0.29\linewidth]{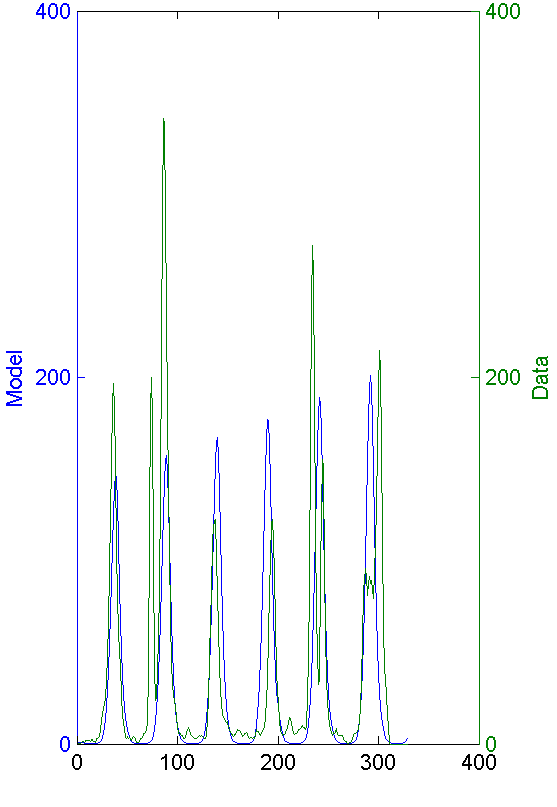}
\hfill (b)  \includegraphics[width=0.29\linewidth]{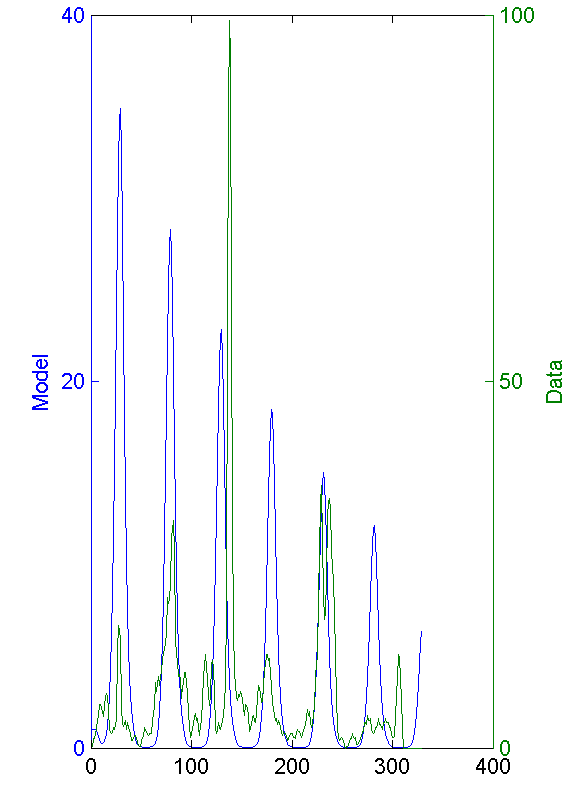}
\hfill (c)  \includegraphics[width=0.29\linewidth]{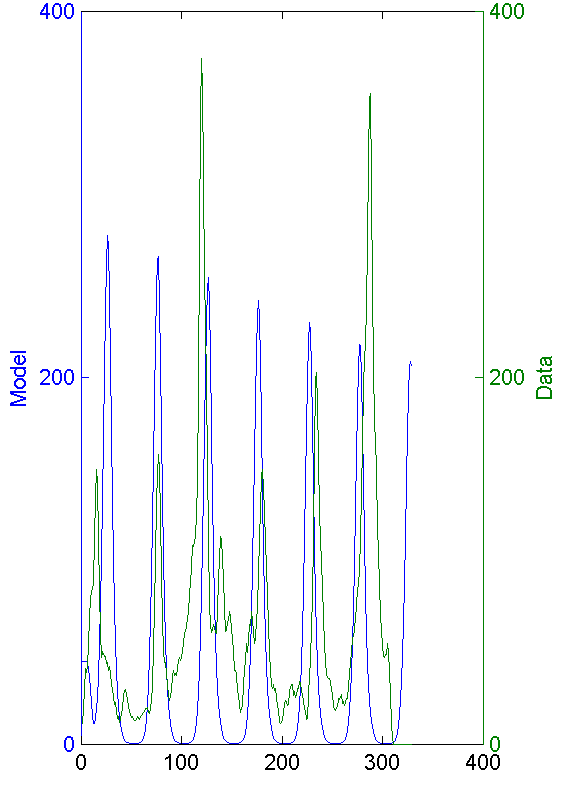}
\caption{The best fits from the non-linked model for a) patch 1, b) patch 2, and c) patch 3. }
\label{fig:betas} 
 \end{figure}

We set up a three patch model with no inter-patch transport in order to calculate the transmission parameters for each patch. We use least squares and Pearson chi-squared statistics to fit the model predictions to the epidemic data for each of the patches. We assume $\beta_{vi} (t)$ is a function of the form $\beta_{vi}(t) = \beta_0 + \epsilon \sin(\frac{2 \pi t + \phi}{365})$. We perform 10000 iterations in which we draw $\beta_0$ and $\epsilon$ at random from a uniform(0,1) distribution and we assess the model fit with these parameters. The ranges of the top 5 coefficients for each of the regions can be found in Table \ref{table:betafitting}, and the top model predictions are shown overlaid with the data in Figure \ref{fig:betas}. We also experiment with assigning random values between 0 and 365 days for the offset $\phi$. The best fit values are centered around 0 as expected -- the data starts in January, which is the onset of autumn in Peru so the disease transmission should be rising early in the data set. The seasonality parameter $\epsilon$ varies more in the coast than the jungle (as expected), however in some of the coastal models is it quite low (0.05), which is somewhat surprising. 

\begin{table}[!htbp]
\label{table:betafitting}
\small
\centering
\begin{tabular}{lllll}
Patch &$\beta_{vi}$ range& $\beta_{hi}$ range & $\phi$ range & $\epsilon$ range\\
\hline
1 & 0.25-0.35 & 0.3-0.4& 291-61 & 0.05-0.43\\
2& 0.3-0.35 & 0.25-0.3 & 296-59& 0.10-0.48 \\ 
3& 0.2-0.4 & 0.25-0.45& 335-103 & 0.14-0.19\\
\end{tabular}
\caption{The parameters fitting the climate data for each of the three patches. }
\label{table:betafitting}
\end{table}

\subsection{Gravity three-patch model: data fitting}

We attempt to fit a three-patch metapopulation model with link weights between the patches predicted by a gravity model, and study the connectivity structure.  In our case, as we do not have transportation data, we will attempt to determine these parameters by data fitting to the disease incidence data.  

We first fix $ \alpha = \gamma = 1$ and fit over 10000 repeats to obtain values for  $\beta$ and $\theta$. We then use these values in another fitting, where we search for optimal values of  $\alpha$ and $ \gamma$. 

\begin{figure} [h!tbp]
\centering
\includegraphics[width=0.6\linewidth]{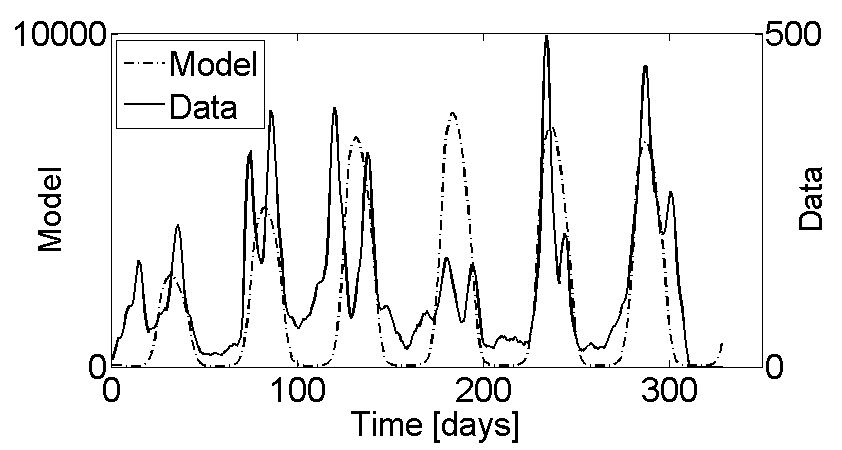}
\caption{The best fit from the three patch model. }
\label{fig:3patch-fullfit}
 \end{figure}

The  best fit we obtained using the least squares and the Pearson chi-squared method picks the yearly peak locations well, but it overestimates the number of infected individuals by a factor of 20  (see Figure \ref{fig:3patch-fullfit}). The parameter values for this run were: $\alpha = 7.53 \times 10^{-9}, \beta = 7.88 \times 10^{-7}, \gamma = 4.2 \times 10^{-7}, \theta =7.17\times10^{-10}, \beta_{vi} = 0.40, \beta_{hi} = 0.03, \phi = 160, \epsilon = 0.1882$. We note that the value for $ \beta_{hi} = 0.03$ is much smaller than the one estimated in single patches, the gravity model parameters are very small and the offset $\phi$ is almost the opposite to our earlier estimate of 0. 

\subsection{Results}

The structure of the Peruvian dengue data is quite complicated due to the seasonality and irregularity of the disease occurrences, and the yearly variation in epidemic size. Additionally,  this simple model does not account for is the complexity of the immunological landscape, with many individuals immune to more than one dengue strain. 

Fitting for the transmission parameter in single non-linked patches showed that finding peak locations is possible, but emphasized the difficulty in predicting the correct epidemic sizes. Fitting a multipatch model to real disease data is more complicated due to having to estimate the contact parameters. Using a gravity model adds for link weights adds four additional parameters that need to be fitted. Despite that, we obtained a good qualitative fit of peak locations. However, the multipatch fits again show a much larger epidemic size than the data. 

We also noticed that the best fit had relatively small gravity model parameters. This suggests that the three-patch structure we were forced to reduce to might not be large enough to properly study the impact of using a gravity model for the link structure. The strength of the gravity model is when there is a non-trivial spatial structure between the locations, and/or their populations are quite different. The starting idea of 7 patches would have been quite interesting, but having to reduce it to a 3-patch model made it less fitting for testing the influence of the gravity model. We will further consider a simpler, nonseasonal scenario in which we attempt to fit a larger number of patches. Additionally, this might be an interesting concept to explore in more depth in the synthetic data setting. 

\section{The 2000-2001 epidemic}
\begin{figure} [h!tbp]
\centering
  \includegraphics[width=0.6\linewidth]{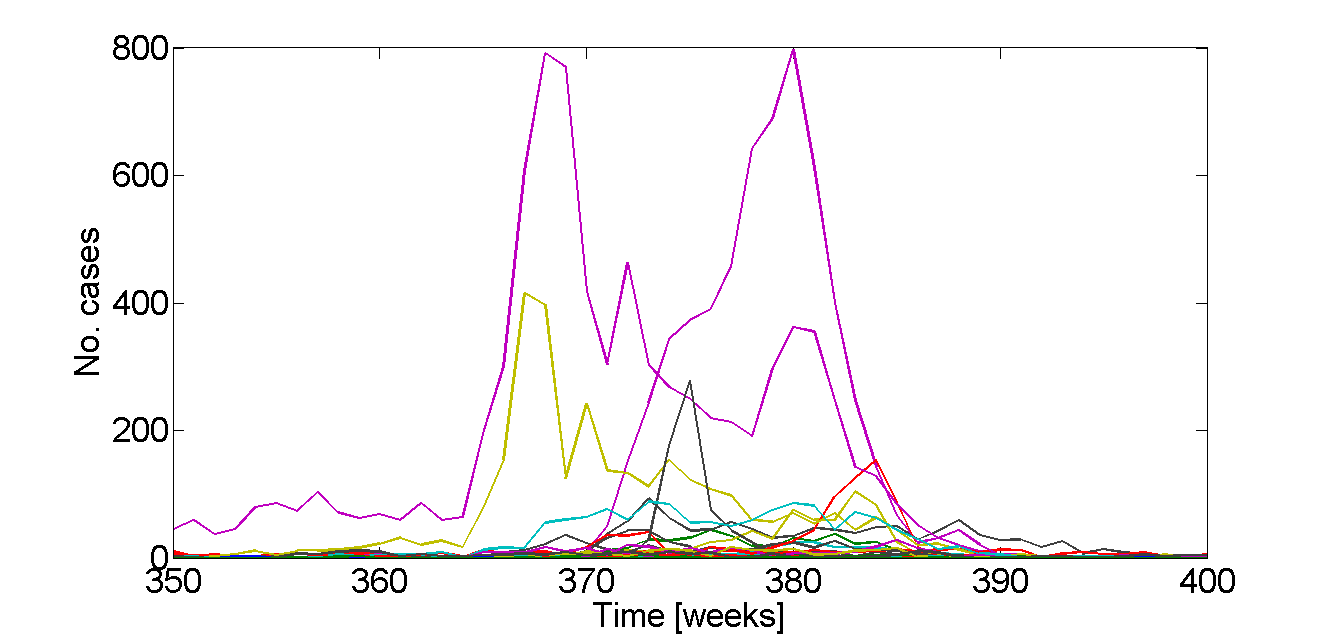}
\caption{The weekly number of cases in each region in the 2000-2001 epidemic. }
\label{fig:dengue2001} 
 \end{figure}

We study a simpler, non-seasonal model by focusing on the data from the 2000-2001 epidemic (weeks 350-400 of the data set). Because the majority of provinces experienced dengue cases in a similar temporal pattern during that epidemic (a typical epidemic curve -- see  Figure \ref{fig:dengue2001}), and the seasonality plays only a small effect in the short-term epidemic, we are able to treat this epidemic as a simple non-seasonal model. We will present  both a three patch model and a 49 patch model (all provinces that experienced the epidemic). 

\subsection{The model}
 The simplified non-seasonal model for the 2000-2001 epidemic is:
\begin{align} \nonumber
\dot{S}_{vi} &= \mu_{vi} N_{vi} - \beta_{vi} \left (\sum_{i=1}^{n}{P_{ij} \frac{I_{hi}}{N_{hi}}} \right )S_{vi} - \mu_{vi}S_{vi}\\ \nonumber
\dot{E}_{vi} &= \beta_{vi} \left (\sum_{i=1}^{n}{P_{ij} \frac{I_{hi}}{N_{hi}}} \right )S_{vi} - \mu_{vi}E_{vi} - \kappa E_{vi}\\ \nonumber
\dot{I}_{vi} &=  \kappa E_{vi} - \mu_{vi}I_{vi}\\ \nonumber
\dot{S}_{hi} &= \mu_{hi} N_{hi} - \beta_{hi} P_{ii} \frac{I_{vi}}{N_{vi}} S_{hi} - \mu_{hi}S_{hi}\\ \nonumber
\dot{E}_{hi} &=  \beta_{hi} P_{ii} \frac{I_{vi}}{N_{vi}} S_{hi} -\lambda E_{hi} - \mu_{hi}E_{hi}\\ \nonumber
\dot{I}_{hi} &= \lambda E_{hi} -\delta I_{hi} - \mu_{hi}I_{hi}\\ \nonumber
\dot{R}_{hi} &= \delta I_{hi} - \mu_{hi}R_{hi},\\ \nonumber
\end{align}
where $P_{ij} = A \frac{n_i^\alpha n_j^\beta}{d_{ij}^\gamma}$.

We use the same parameter values as before (see Table \ref{table:modelparameters}), but remove the seasonality.

\subsection{Three patch model}
\begin{figure} [h!tbp]
\centering
 \hfill (a) \includegraphics[width=0.28\linewidth]{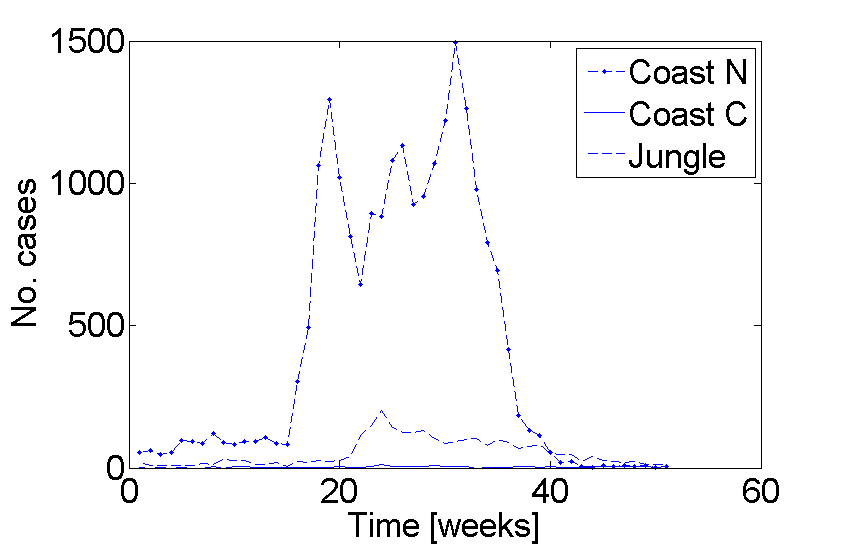}
 \hfill (b) \includegraphics[width=0.28\linewidth]{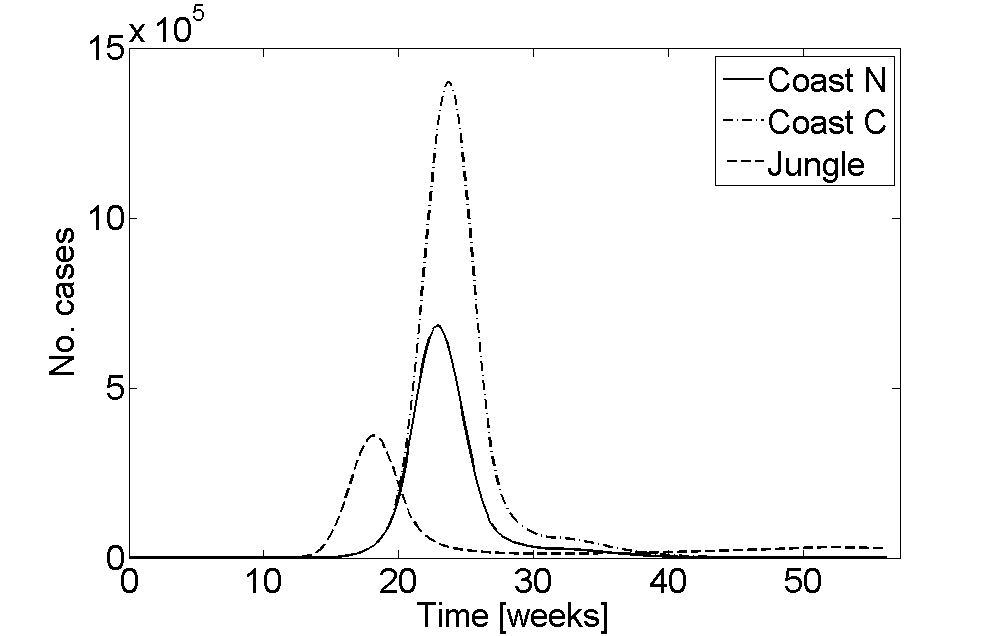}
 \hfill (c) \includegraphics[width=0.28\linewidth]{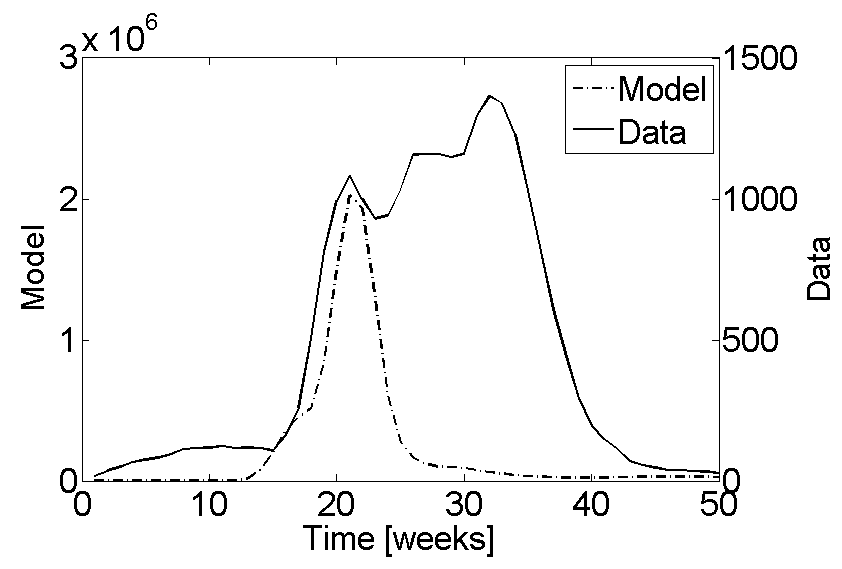}
\caption{The best fitting of the 1000 runs of the 2000-2001 epidemic fitting. (a) The actual case numbers for the three patches and (b) the model predictions and (c) data and the sum of model predictions side by side. }
\label{fig:3patch-20001} 
 \end{figure}

Here we fit the gravity model parameters for the same three patch structure as before, using the 2000-2001 data and the same parameters as obtained via best fits earlier. We observe that in the model the central coast peaks first, and the north coast is delayed by about 10 weeks, giving the wider peak in the summed countrywide data. Since most of the peaks in the dengue data are wide and formed from multiple provincial peaks, this is in agreement with the data (see Figure \ref{fig:3patch-20001}a). This supports the notion that ideally the model should be fitted to smaller patches. On the country level, we notice the location and shape of the summed peak fit relatively well, but the size of the epidemic is much larger than in reality (see Figure \ref{fig:3patch-20001}b). 

\subsection {49 patch model}
We also attempt to fit a 49 patch model --  a more complicated spatial structure, which gives more room for the weights predicted by the gravity model to differ form a simple arbitrary structure. This might be able to give more variation in the timing of local epidemics, causing wider peaks as observed in the real data.

\subsubsection{Gravity model}

\begin{figure} [h!tbp]
\centering
 \hfill (a) \includegraphics[width=0.45\linewidth]{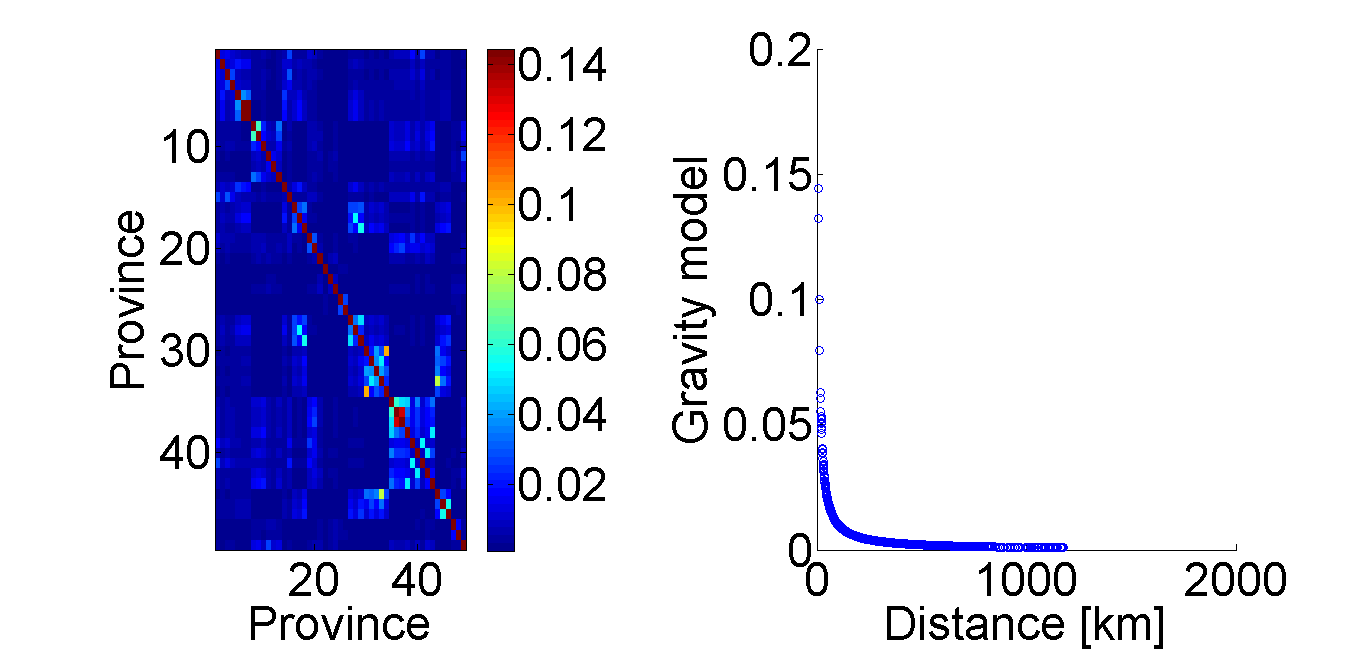}
 \hfill (b) \includegraphics[width=0.45\linewidth]{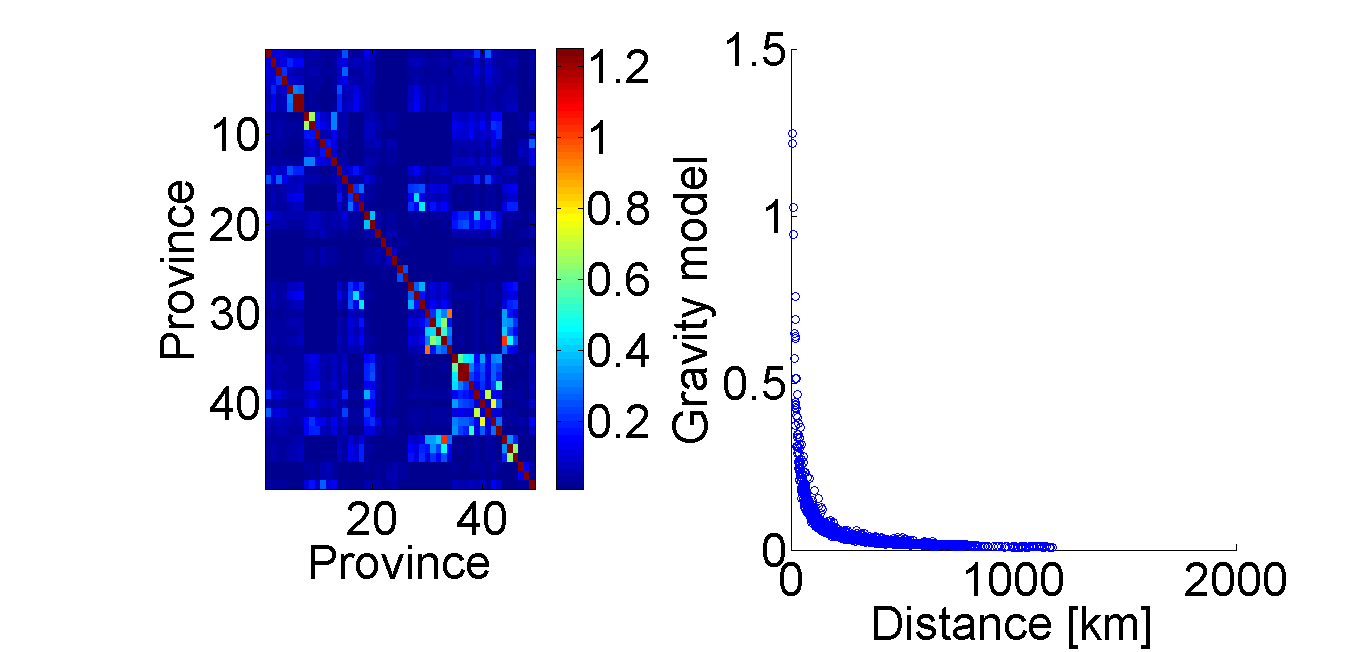}\\
\hfill \\
 \hfill (c) \includegraphics[width=0.45\linewidth]{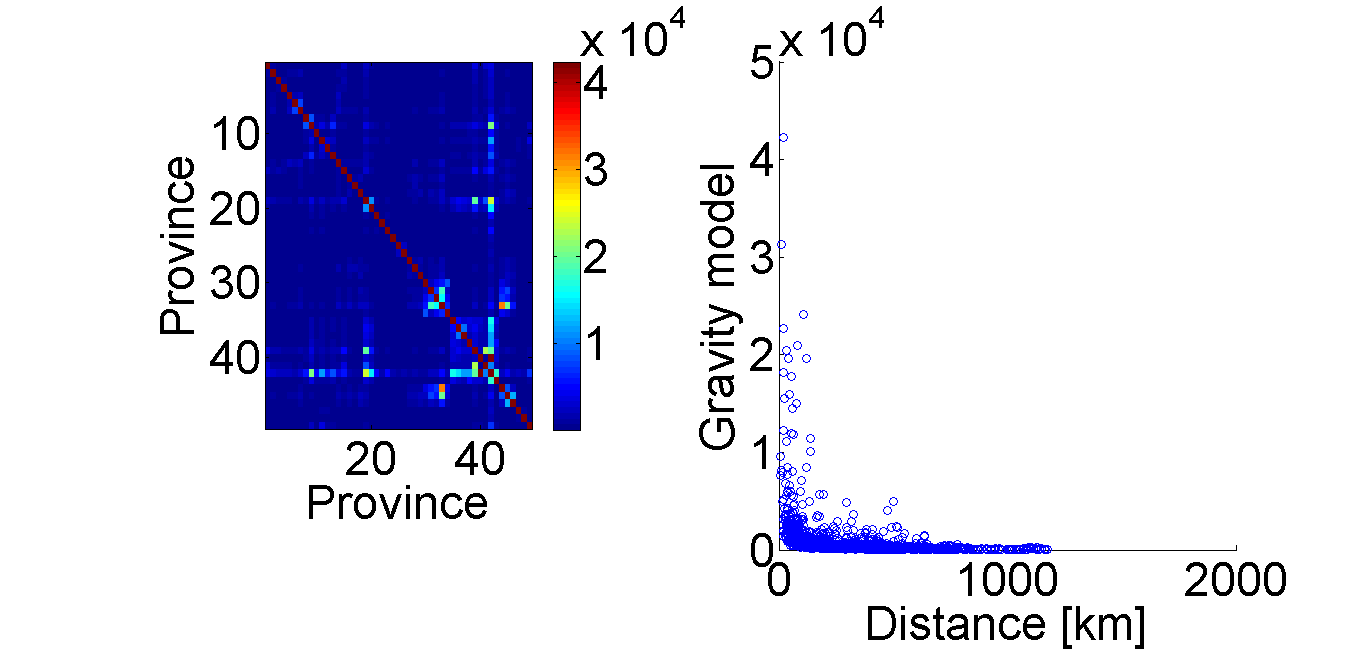}
 \hfill (d) \includegraphics[width=0.45\linewidth]{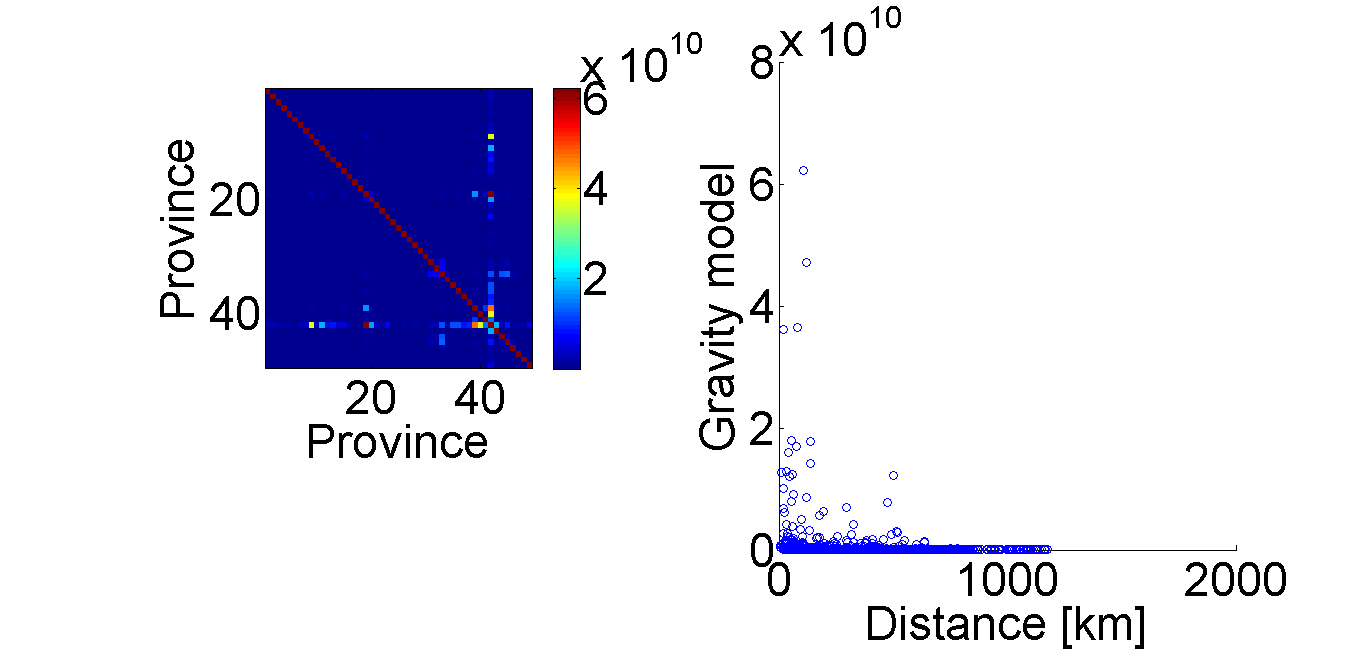}\\
\caption{The effect of the $\alpha$ and $\beta$ parameters on the gravity model predictions: (left) the province-province interaction predictions and (right) scatter plot of distance vs gravity model predictions. at $\gamma = 2, \theta = 1$ and (a) $\alpha = \beta = 0.001$,  (b) $\alpha = \beta = 0.1$, (c) $\alpha = \beta = 0.5$, (d) $\alpha = \beta = 1$. }
\label{fig:grav1-ab} 
 \end{figure}

In this section we present the predictions form a gravity model for link weights when varying the gravity model parameters. Figure \ref{fig:grav1-ab} shows that with increase in the province population exponents $\alpha$ and $\beta$ more importance is given to links with the most populous patches (here: Lima) and the gravity model prediction curve against distance flattens.

\begin{figure} [h!tbp]
\centering
\hfill \\
 \hfill (a) \includegraphics[width=0.45\linewidth]{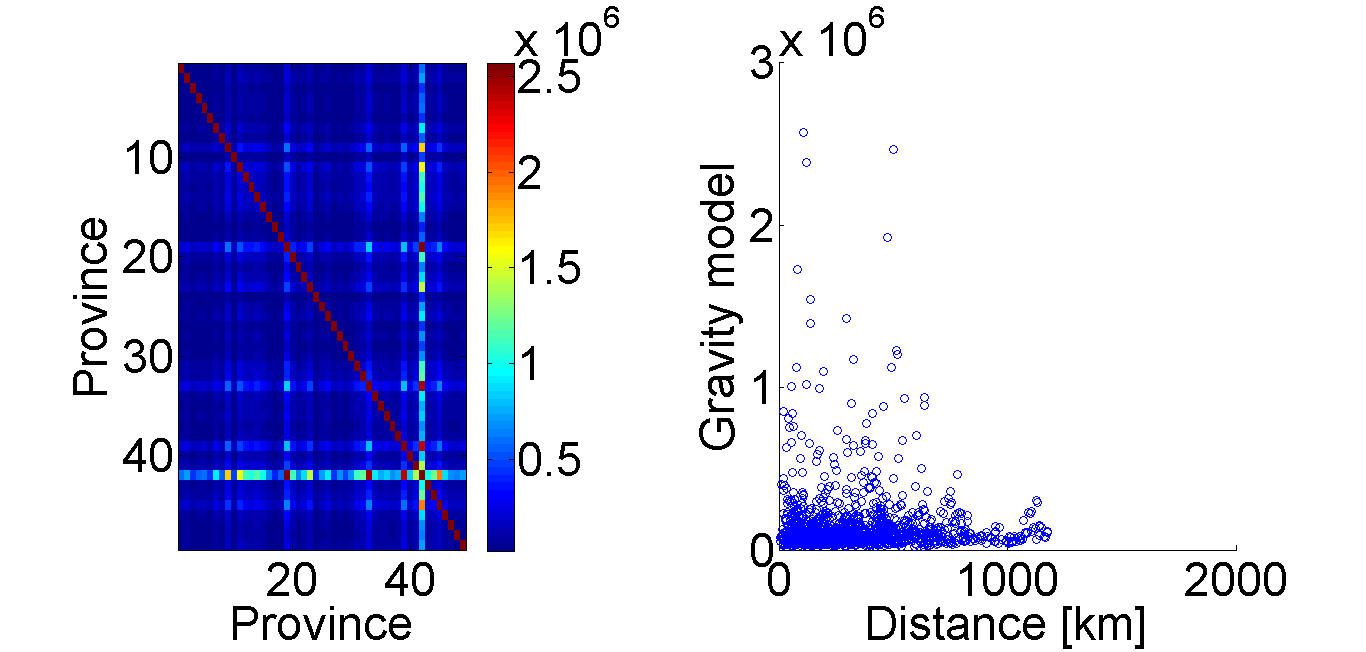}
 \hfill (b) \includegraphics[width=0.45\linewidth]{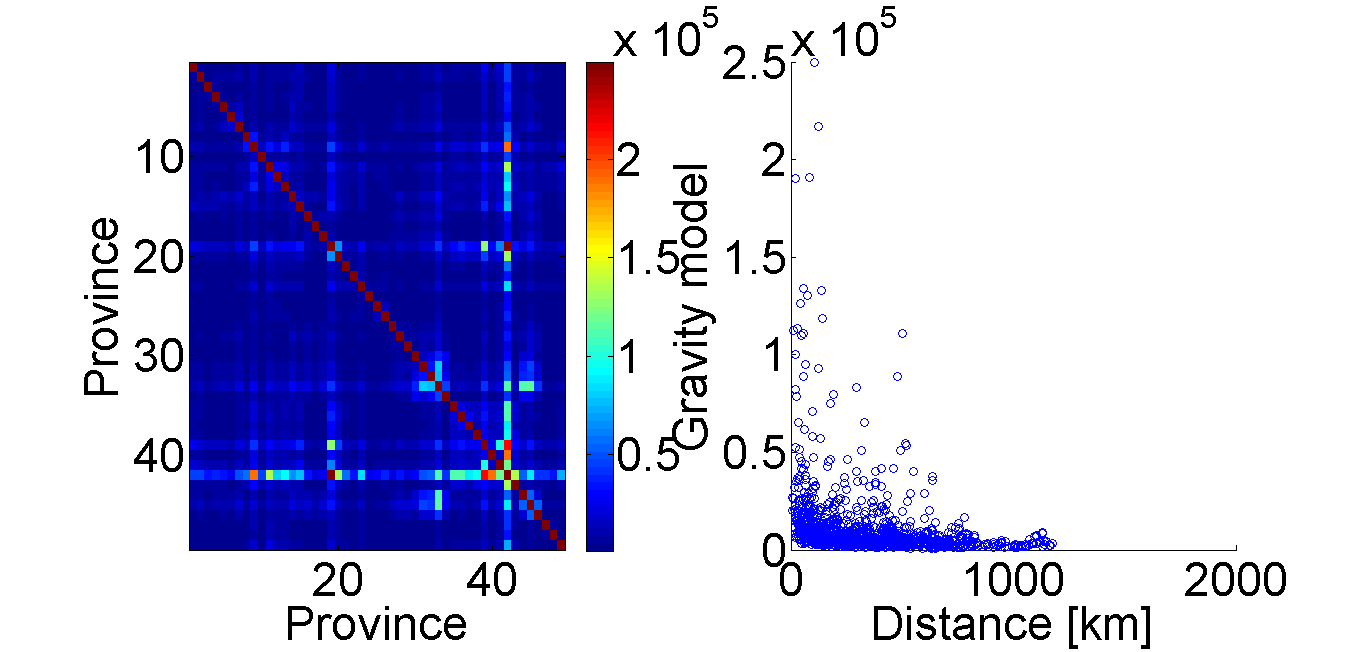}
\hfill \\
 \hfill (c) \includegraphics[width=0.45\linewidth]{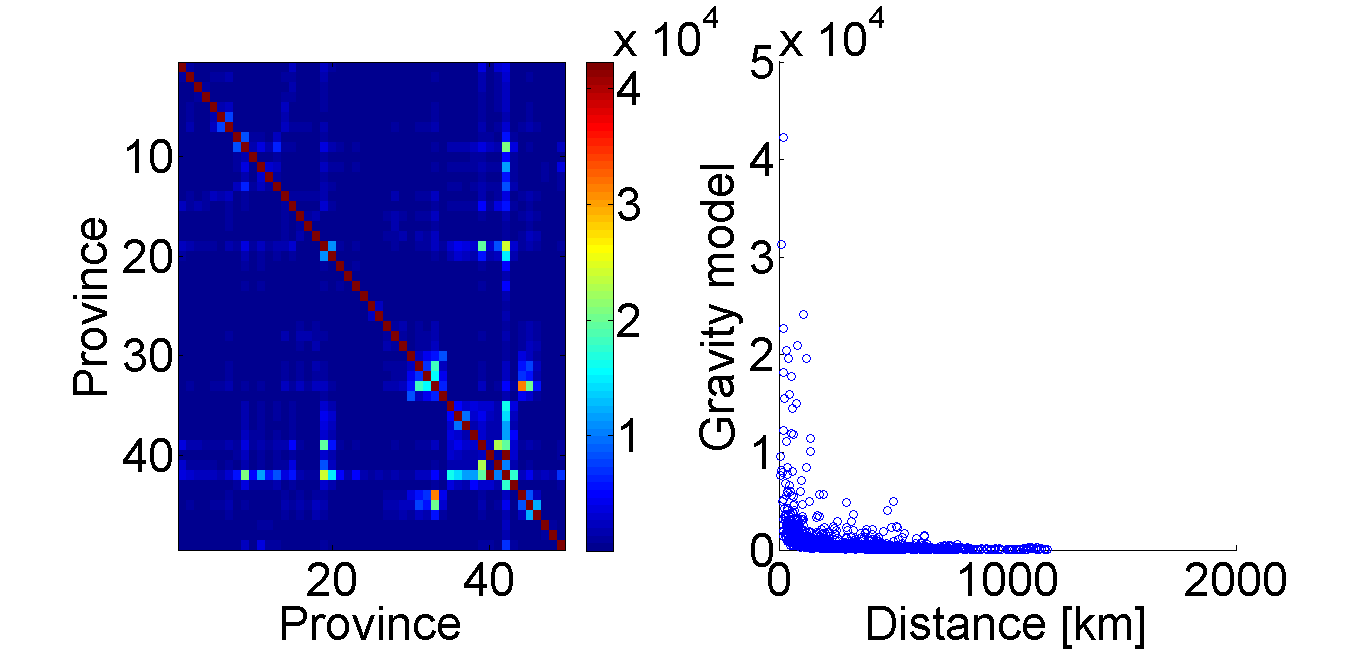}
 \hfill (d) \includegraphics[width=0.45\linewidth]{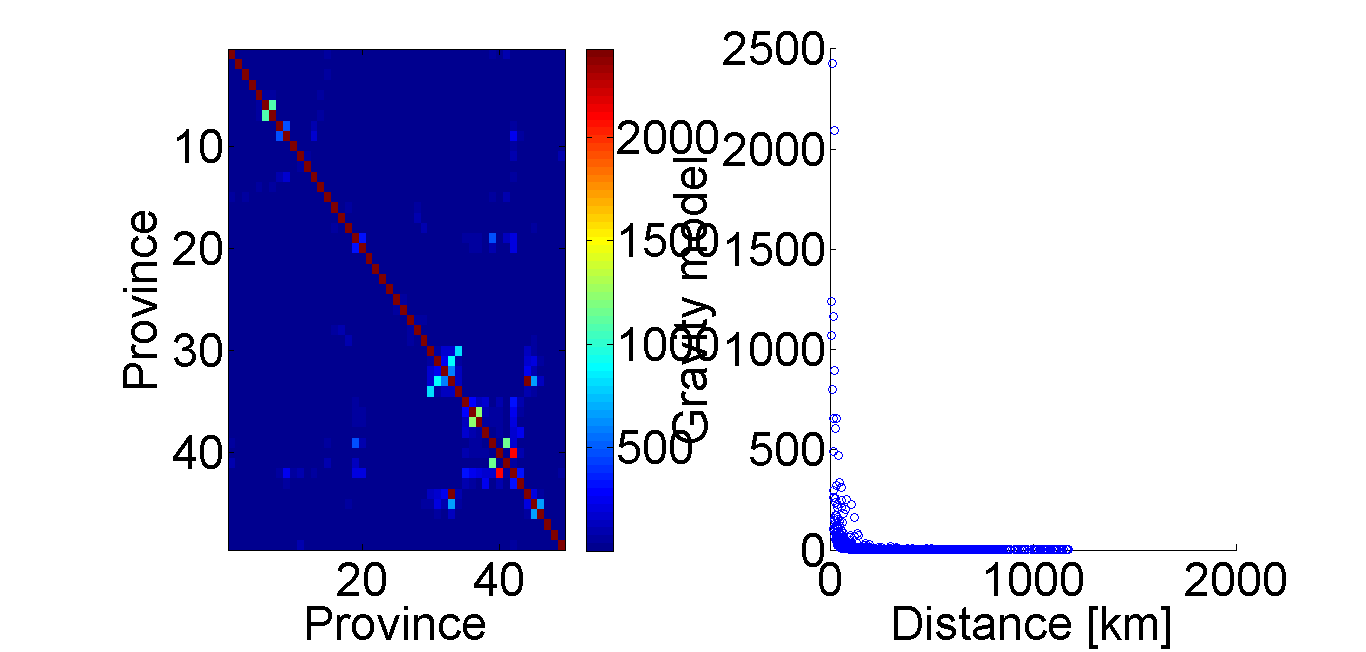}\\
\caption{The effect of the $\gamma$ parameter on the gravity model predictions: (left) the province-province interaction predictions and (right) scatter plot of distance vs gravity model predictions. at $\alpha = 1,\beta = 1, \theta = 1$ and (a) $ \gamma = 0.001$,  (b) $ \gamma = 0.5$,  (c) $ \gamma = 1$,  (d) $ \gamma = 2$. }
\label{fig:grav1-g} 
 \end{figure}

Figure \ref{fig:grav1-g} shows the changes in gravity model predictions as we vary the distance exponent $\gamma$. The shape of the gravity model prediction against distance becomes more curved as $\gamma$ increases, although the change is less obvious than in the synthetic setting. As $\gamma$ increases the gravity model prediction becomes dominated by connections between close-by, highly populated provinces. 

The gravity model predictions are based on the large populations, and as a result they often exceed 1 and thus need scaling to a level suitable to be used for link weights using the parameter $\theta$. All the gravity model parameters are fitted from data using least squares and Pearson chi-squared statistic.

\subsubsection{Data fitting}

\begin{figure} [h!tbp]
\centering
\hfill \\
 \hfill (a) \includegraphics[width=0.45\linewidth]{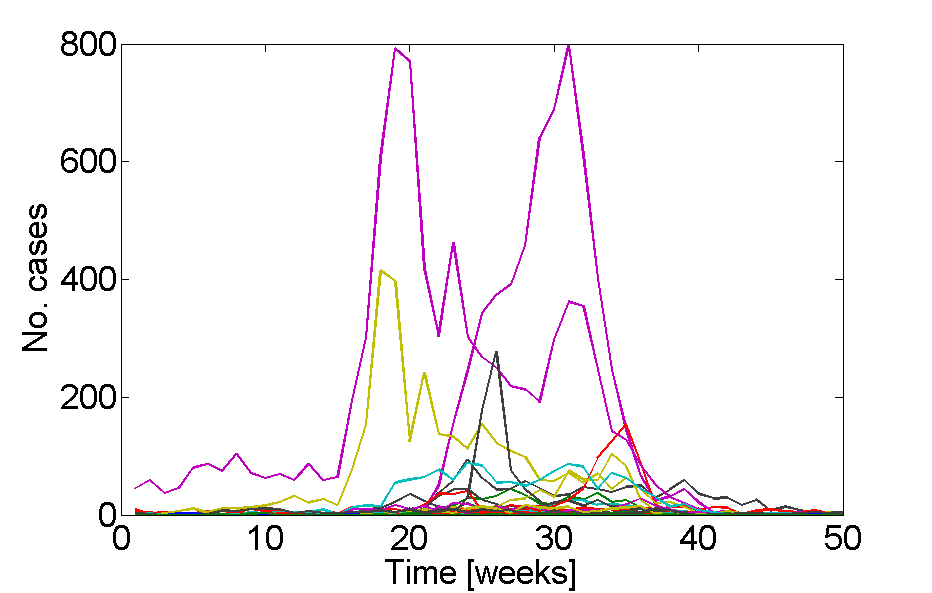}
 \hfill (b) \includegraphics[width=0.45\linewidth]{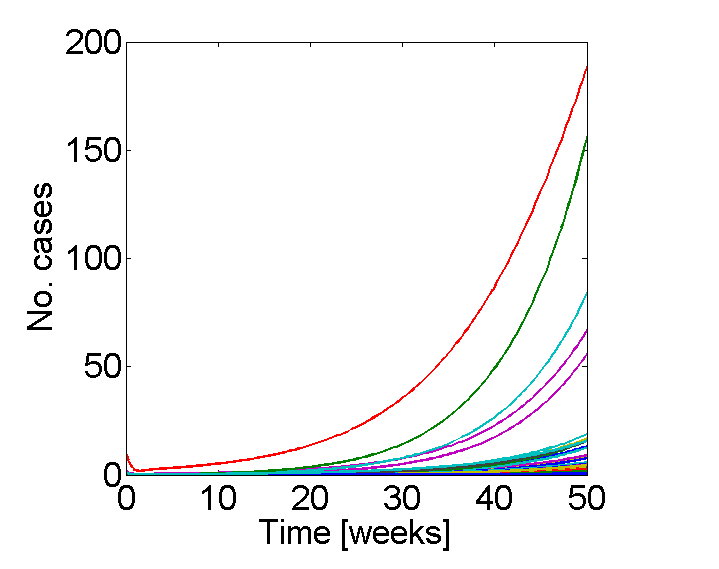}\\
 \hfill (c) \includegraphics[width=0.45\linewidth]{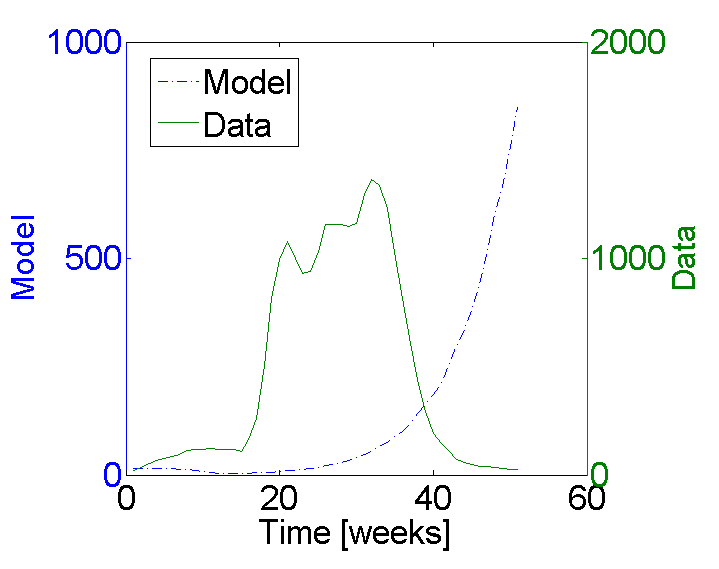}
 \hfill (d) \includegraphics[width=0.45\linewidth]{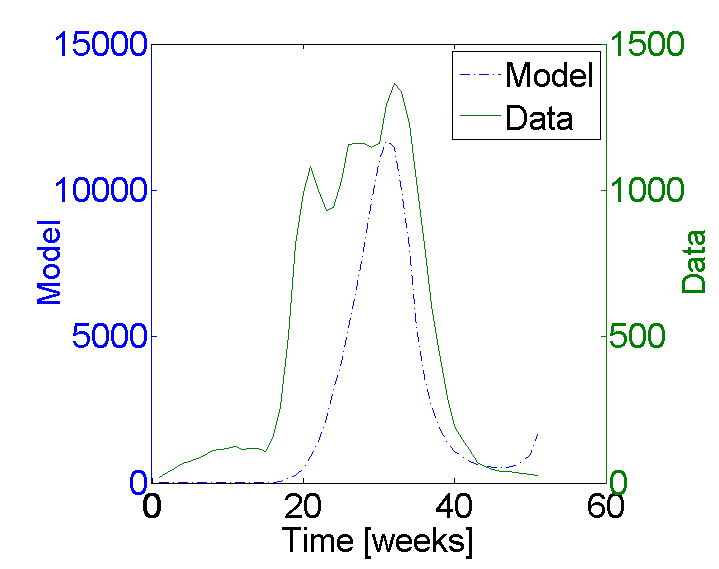}

\caption{The results of the fitting to the 49 patch model: (a) the number of dengue cases, (b) the results of the 49 patch models, (c) the data and best fitting model on the same plot, (d) an example of a model that is better shaped but far off in values. }
\label{fig:49patches} 
 \end{figure}

We run 10000 repeats of fitting the model to the data. Each time, we select $\alpha, \beta, \gamma$ and $\theta$ at random. At each run, we take a random value between 0 and 1 and multiply it by a randomly selected multiple of 10 from $(10^{-10}, 10^{-9}, \ldots, 1)$. In this way we achieve a spread over the parameter space with a focus on small parameter values (from previous experience, these are expected to be better fitting). The best fits can be seen in Figure \ref{fig:49patches}. We notice that the fitting is not of a similar shape to the epidemic curve. When looking into the different results, we notice certain parameter sets generated a similar epidemic curve to the real data, however the numbers of infected people were much higher (up to $10^5$) -- this scores worse in the least squares and Pearson chi-squared fitting than the similar in numbers, yet different in shape result. This suggests maybe another scoring function would perform better in selecting parameter values in this setting.

The parameters for the result in Figure  \ref{fig:49patches}d were: $\alpha = 1.82\times 10^{-8}, \beta = 3.56\times 10^{-5}, \gamma = 0.86, \theta = 0.30, \beta_{vi} = 0.23, \beta_{hi} = 0.228747119741029$. We notice that $ \beta $ and $\gamma$ are larger this time, potentially suggesting the dependence on distance is more important in the more detailed 49 patch model. 

\section{Discussion}
Studies have shown that in case of diseases that are spread by human movement, in contrast to animal-born diseases, reaction-diffusion is not a god model for spatial spread \cite{Xia2004}. Gravity models have been considered as an alternative method of spread, which is able to take account of the propensity to travel along the fastest/simplest route, and factors such as population/importance of cities that influence human transportation patterns  \cite{Xia2004}. 

We have shown that the gravity model has the potential to be used to predict link weights for a multipatch model in a synthetic setting. It gives the model the potential to have a not-fully synchronized set of follower patches driven by one driving patch. The degree of correlation between the driver and followers depends on the distance between them and their populations. However, fitting a gravity model to real data is a complicated process, requires very good data quality, spatial and temporal resolution, and time.  

The first study used dengue data from 2002-2008 and it focused on seasonal dengue transmission, which shows yearly extinctions of the disease on the coast and yearly reinfections from the endemic jungle regions. We were forced to reduce model complexity from an initial plan of 7 patches to 3 climate-based patches due to poor data quality, irregular occurrence of dengue in provinces making seasonal modeling difficult for this particular setting. Additionally, the climate data was not fully available for all the patches. We found that while optimizing parameters for each patch on its own worked relatively well, optimizing the parameters for the multipatch model with gravity connections proved difficult, and even with the multiple aggregations, we found the data highly irregular and difficult to fit. We still managed to find peaks that collocated with peaks in the data, however the size of the epidemic was too large in the predictions. Perhaps, more climate data were available, a model that incorporates the climate data explicitly into the transmission parameter might be better performing, as it would link the temperature factor to presence or lack of outbreaks. Additionally, the predictions could potentially be improved by having a stochastic model, in which case we would be able to better model the periodic extinctions and reinfections often experienced by many regions with regard to dengue fever. 

We then decided to simplify the model by removing the effect of seasonality and studying only the 2000-2001 countrywide dengue epidemic. We showed both a 3 patch model with climate-based patches, and a 49 patch model representing all of the provinces that experienced the epidemic. The three patch model predicted curves similar in shape to the real data, but more narrow, and with much higher numbers of infected individuals. The 49 patch model experienced the same problems and the best-shaped fit was masked by a different shaped prediction that was scoring better because of being lower in numbers. 

Xia et al. \cite{Xia2004} found that epidemic spread scales inversely with distance (rather then distance squared as would be the case for diffusion), and the population size of epidemic donor cites is important to disease spread. In contrast, in the three patch model, we found the gravity model parameters and predictions to be very small, suggesting that either the transportation structure is not important to the disease spread, or the three-patch scale of the model is not detailed enough for the effects to show. The higher $\beta$ and $\gamma$ values obtained in the preliminary 49-patch fitting appear to support this hypothesis. This question could be further studied in a synthetic setting.

This model is able to capture the yearly peaks of dengue fever case numbers. However, it does not account for all the features relevant to the recurring epidemics. It only models one dengue strain, while there are now four strains in continuous circulation across Peru. Due to the fact that infection with one strain gives immunity to it (but not the others), the apparent yearly epidemics are often caused by different strains in different years. Accounting for more than one strain would improve the accuracy of the model. 

One potential extension of this study that would be very useful would be to include transportation data to compare the gravity model predictions to and potentially fit the gravity model parameters $\alpha$, $\beta$ and $\gamma$ to the transport data. 

Another extension of the model, if only on a theoretical level, could be to see the differences between the disease patterns in a system using a gravity model and the recently proposed alternative, the radiation model \cite{Simini2012}. This model is able to better predict interactions in crowded spaces, where the population density in cities between two provinces of interest might increase (if high) or decrease (if low) the disease spread compared to gravity spread that would be predicted based on distance and population only. 


%
%

\section*{Acknowledgments} 
We would like to thank Dr.~Carlos Castillo-Chavez, Executive Director of the Mathematical and Theoretical Biology Institute (MTBI), for giving us the opportunity to participate in this research program.  We would also like to thank Co-Executive Summer Directors Dr.~Erika T.~Camacho and Dr.~Stephen Wirkus for their efforts in planning and executing the day to day activities of MTBI. We also want to give special thanks to Dr. Sherry Towers for her help in obtaining data and defining the model, and Dr.~Carlos Castillo-Chavez again, for extremely helpful comments and reviewing the paper. This research was conducted in MTBI at the Mathematical, Computational and Modeling Sciences Center (MCMSC) at Arizona State University (ASU). This project has been partially supported by grants from the National Science Foundation (NSF - Grant DMPS-1263374), the National Security Agency (NSA - Grant H98230-13-1-0261), the Office of the President of ASU, and the Office of the Provost of ASU.

\bibliography{dengugraphbiblio}

\end{document}